\documentclass[fleqn,usenatbib]{mnras}

\usepackage{newtxtext,newtxmath}
\usepackage{enumitem}

\usepackage[T1]{fontenc}

\DeclareRobustCommand{\VAN}[3]{#2}
\let\VANthebibliography\thebibliography
\def\thebibliography{\DeclareRobustCommand{\VAN}[3]{##3}\VANthebibliography}

\usepackage{graphicx}	
\usepackage{amsmath}	
\usepackage{caption}
\usepackage{subcaption}

\newcommand{\Ha}{H$\alpha$}
\newcommand{\OIII}{[{O}~{\scriptsize {III}}]}
\newcommand{\OII}{[{O}~{\scriptsize {II}}]}
\newcommand{\Hb}{H$\beta$}

\title[CSST Large-scale Structure Analysis Pipeline: II.]{CSST Large-scale Structure Analysis Pipeline: II. the CSST Emulator for Slitless Spectroscopy (CESS)\thanks{https://github.com/RainW7/CSST-grism-emulator/}}

\author[R. Wen et al.]{Run Wen,$^{1,2}$
Xian~Zhong Zheng,$^{1,2}$\thanks{E-mail: xzzheng@pmo.ac.cn}
Yunkun Han,$^{3}$
Xiaohu Yang,$^{4,5}$
Xin Wang,$^{6,7}$
Hu Zou,$^{7}$
\newauthor{Fengshan Liu,$^{6,7,8}$
Xin Zhang,$^{7,8}$
Ying Zu,$^{4}$
Dong Dong Shi,$^{1}$
Yizhou Gu$^{4}$
and Yirong Wang$^{4}$}
\\
$^{1}$Purple Mountain Observatory, Chinese Academy of Sciences, 10 Yuanhua Road, Nanjing 210023, China\\
$^{2}$School of Astronomy and Space Science, University of Science and Technology of China, Hefei 230026, China\\
$^{3}$Yunnan Observatories, Chinese Academy of Sciences, 396 Yangfangwang, Guandu District, Kunming 650216, China\\
$^{4}$Department of Astronomy, School of Physics and Astronomy, Shanghai Jiao Tong University, Shanghai 200240, China\\
$^{5}$Tsung-Dao Lee Institute and Key Laboratory for Particle Physics, Astrophysics and Cosmology, Ministry of Education, Shanghai 201210, China\\
$^{6}$School of Astronomy and Space Science, University of Chinese Academy of Sciences (UCAS), Beijing 100049, China\\
$^{7}$National Astronomical Observatories, Chinese Academy of Sciences, 20A Datun Road, Chaoyang District, Beijing 100101, China\\
$^{8}$Key Laboratory of Optical Astronomy, National Astronomical Observatories, Chinese Academy of Sciences, 20A Datun Road, Chaoyang District, \\Beijing 100101, China
}

\date{Accepted 2024 January 8. Received 2023 December 27; in original form 2023 October 27}

\pubyear{2024}

\begin{document}
\label{firstpage}
\pagerange{\pageref{firstpage}--\pageref{lastpage}}
\maketitle

\begin{abstract}
The Chinese Space Station Telescope (CSST) slitless spectroscopic survey will observe objects to a limiting magnitude of $\sim23$\,mag (5$\sigma$, point sources) in $U$, $V$, and $I$ over 17\,500\,deg$^2$.  The spectroscopic observations are expected to be highly efficient and complete for mapping galaxies over $0<z<1$ with secure redshift measurements at spectral resolutions of $R\sim200$, providing unprecedented data sets for cosmological studies.  To quantitatively examine the survey potential, we develop a software tool, namely the CSST Emulator for Slitless Spectroscopy (\texttt{CESS}), to quickly generate simulated 1D slitless spectra with limited computing resources.  We introduce the architecture of \texttt{CESS} and the detailed process of creating simulated CSST slitless spectra.  The extended light distribution of a galaxy induces the self-broadening effect on the 1D slitless spectrum.  We quantify the effect using morphological parameters: S\'ersic index, effective radius, position angle, and axis ratio.  Moreover, we also develop a module for \texttt{CESS} to estimate the overlap contamination rate for CSST grating observations of galaxies in galaxy clusters.  Applying \texttt{CESS} to the high-resolution model spectra of a sample of $\sim140$ million galaxies with $m_{z}<21$\,mag selected from the Dark Energy Spectroscopic Instrument LS DR9 catalogue, we obtain the simulated CSST slitless spectra.  We examine the dependence of measurement errors on different types of galaxies due to instrumental and observational effects and quantitatively investigate the redshift completeness for different environments out to $z\sim1$.  Our results show that the CSST spectroscopy is able to provide secure redshifts for about one-quarter of the sample galaxies. 
\end{abstract}




\section{Introduction}

Cosmology is an essential frontier branch of astrophysics that studies the origin and evolution of the Universe.  By analysing the large-scale structure of galaxies to measure baryon acoustic oscillations, redshift-space distortions, and large-scale galaxy clustering, researchers can extract crucial insights into the expansion history of the Universe, the formation and evolution of cosmic structures, the nature of dark matter and dark energy.  Together, the two components account for approximately 95\,per\,cent of the Universe's composition.

The imaging and spectroscopic surveys of galaxies serve as fundamental observational tools for cosmological research.  In the past couple of decades, large-scale surveys of this kind, including the 2dF Galaxy Redshift Survey \citep{Peacock+2001,Percival+2001}, the Sloan Digital Sky Survey \citep{Tegmark+2004,Alam+2017}, the Kilo-Degree Survey \citep{deJong+2015,SpurioMancini+2019} and the Dark Energy Survey \citep{DESC+2016, Abbott+2018}, have provided large amounts of data for mapping the three-dimensional (3D) distribution of cosmic large-scale structures.  The so-called Stage~IV cosmic surveys based on the current and upcoming ground- and space-borne telescopes will continuously map the Universe with larger survey areas and higher sensitivities \citep{Weinberg+2013}.  The currently underway facilities dedicated to these Stage~IV surveys include the Dark Energy Spectroscopic Instrument \citep[DESI; ][]{Dey+2019} and the \textit{Euclid} \citep{Laureijs+2011}.  In the near future, projects such as the Rubin Observatory Legacy Survey of Space and Time \citep[LSST; ][]{LSC+2009}, the \textit{Chinese Space Station Telescope} \citep[CSST; ][]{Zhan+2011}, and the \textit{Roman Space Telescope} \citep[formerly known as WFIRST; ][]{Spergel+2015}, will conduct observations, providing us with observational data of hundreds of millions of galaxies.

CSST is a 2-m space telescope with a large field of view (FOV) and high spatial resolution dedicated to wide-field surveys \citep{Zhan+2011}.  It is equipped with six instruments, including the survey camera with 30 9K\,$\times$\,9K CCD detectors in a 5$\times$6 mosaic array covering 1.1\,deg$^2$.  This mission is expected to start its science operation in 2025 and has a lifetime of 10\,yr \citep{Zhan+2021}.  CSST will share the same orbit as the China Manned Space Station and fly independently at a reachable distance.  With the advantages of both large FOV and high-image quality, CSST will be able to carry out in-depth cosmological observations.  One of the main goals of CSST is to perform a wide multiband imaging and slitless spectroscopic survey over 17\,500\,deg$^2$, using in total of 70\,per\,cent of mission lifetime.  The filters and gratings are installed in front of different CCD detectors in the focal plane.  There are six gratings equally divided into three bands, $GU$ (255--400\,nm), $GV$ (400--650\,nm), and $GI$ (620--1000\,nm), providing spectral resolutions of $R\gtrsim 200$ and a total wavelength coverage from 255 to 1000\,nm.  The CSST spectroscopic redshift survey aims to fulfil some valuable cosmological science goals, such as the accelerated expansion of the Universe and the nature of dark energy and dark matter, galaxy clustering, and the large-scale structures of the Universe \citep{Cao+2018,Gong+2019,Zhang+2019,Zhouxc+2021,Miao+2023,Zhang+2023}.

Purity and completeness of galaxy samples are two key measures in exploring the large-scale galaxy clustering through the slitless spectroscopic redshift surveys \citep{Weinberg+2013,Reuter+2020}.  The purity is influenced by the statistics and systematic errors of spectroscopic redshifts, including the outlier rate and the mix-up of emission lines.  In contrast, the 3D distribution of cosmic large-scale structures is extremely sensitive to redshift completeness in line with the radial and lateral variations, which seriously influence the measurement accuracy of galaxy clustering \citep{Weinberg+2013,Alam+2017}.  A detailed analysis of the factors that influence redshift measurements in the survey is thus crucial to determining cosmological parameters.  Therefore, extensive tests on the outputs of the slitless spectroscopic survey are strongly demanded before the mission starts.  Within this framework, a software tool to generate realistic mock grating spectra from the CSST slitless spectroscopic survey will help to commit such tests.

In this work, we present a program named the CSST Emulator for Slitless Spectroscopy (\texttt{CESS}), to quickly generate one-dimensional (1D) spectra from the simulated CSST slitless grating spectroscopic observations by making use of a variety of empirical relations of galaxies.  We use the CSST parameters to quantify the instrumental effects for the grating spectroscopy and convert the input high-resolution galaxy spectral templates into the CSST slitless grating spectra.  For extended sources, the self-blending effect on the spectra is also taken into account.  In crowded fields, some spectra may overlap with each other.  The flux contamination from the neighbouring sources decreases the quality of the target spectra and increases the noises.  We further develop a specific module to estimate the overlap rate and contamination fraction from adjacent sources in the crowded fields using the coordinates, lengths, and widths of simulated CSST slitless grating spectra.  We also apply our \texttt{CESS} to a large sample of galaxies with high-resolution model spectra and present the implications for the CSST redshift survey.

This paper is organized as follows.  
Section~\ref{sec:parameters} describes the general outline of the \texttt{CESS} program.  
Section~\ref{sec:methods} gives the detailed process to generate mock spectra.  
In Section~\ref{sec:modules}, we introduce the modules adopted in the program for estimating the effects of self-blending and overlapping.  The illustrative cases are presented in Section~\ref{sec:applications}.  
Finally, we summarize the features of \texttt{CESS} in Section~\ref{sec:conclusion}.  
Throughout this paper, we adopt the flat Lambda cold dark matter cosmology with $\Omega_\mathrm{M}=0.3$, $\Omega_\mathrm{\Lambda}=0.7$, and H$_0 = 70$\,km\,s$^{-1}$\,Mpc$^{-1}$.



\section{General Outline of The Program}
\label{sec:parameters}

In this section, we present the general outline of the program \texttt{CESS} for the CSST slitless spectroscopic survey.  CSST adopts gratings to disperse light from objects and obtain spectra.  The gratings generate multiple-order spectra for a single object.  The first-order spectrum of the object typically contains the majority of the total energy and is widely used as the target spectrum for scientific research \citep{Kuntschner+2011}.

Moreover, the light from the object is dispersed along with a trace in slitless spectroscopy.  In practice, the trace tends to be slightly curved, causing some parts of the trace to deviate by several pixels away from the dispersion direction \citep{Pirzkal+2016}.  We select an extraction region that covers the two-dimensional (2D) spectrum image to extract the 1D spectrum.  Since this extraction region is much broader than the trace offsets from the dispersion axis, we ignore the offsets in our emulator because they have negligible effects on the 1D spectrum.  Throughout this work, we focus on the analysis of the first-order spectra and ignore the small offset of the spectral trace.

\subsection{Workflow} 
\label{sec:workflow}

\begin{figure*}
\centering
\includegraphics[width=0.98\textwidth, angle=0]{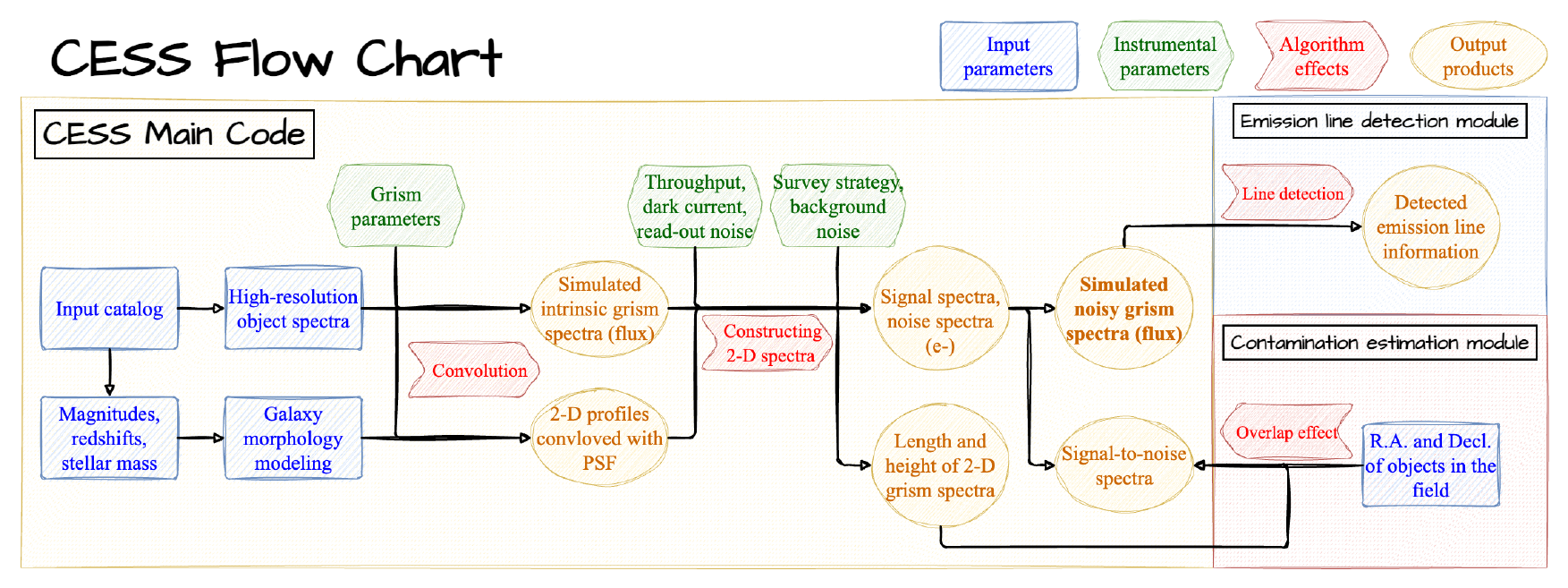}
\caption{ The flow chart of our emulator program \texttt{CESS} for CSST slitless spectroscopic survey.  The blue squares refer to the original input galaxy parameters.  The green hexagons represent the CSST instrumental parameters.  The red arrows stand for the major algorithms used in the program computation.  The brown ellipses show the output products.}
\label{fig:flow_chart}
\end{figure*}

Our program, \texttt{CESS}, is designed to simulate slitless grating spectra for billions of galaxies.  This enables us to generate mock data that can be used for a detailed and quantitative analysis of the potential and systematics for the CSST spectroscopic redshift survey.  It is worth noting that CESS is also capable of simulating slitless spectra for stars.

A visual representation of the workflow of our program is presented in Fig.~\ref{fig:flow_chart}.  This flow chart illustrates the step-by-step process of \texttt{CESS}.  The program utilizes various inputs, including high-resolution flux-calibrated spectra and morphological parameters of a sample of galaxies with known physical parameters of redshift, stellar mass, and magnitude in the optical bands.  The instrumental parameters specific to CSST, such as the total throughput curves and the point spread function (PSF) size in three grating bands, the pixel scale, as well as the dark currents and readout noises of CCD detectors, are collected from mission publications.  The total throughput curves take into account the transmission curves of the telescope optics and grating, as well as the quantum efficiency of the CCD detectors.  In addition, the proposed observing strategies for the key surveys provide essential information for CESS, such as the exposure time and the number of exposures required for each pointing.  Lastly, we adopt the brightness of the sky background observed by the \textit{Hubble Space Telescope} (\textit{HST}) to estimate photons from the unresolved background.

Here, we describe the main steps of the \texttt{CESS} workflow.  First, we downgrade the high-resolution flux-calibrated spectra, which are given in the observed frame, to the CSST grating spectral resolution ($R\gtrsim 200$).  Simultaneously, we use empirical relations of galaxy structures to assign structural parameters to target galaxies and construct two brightness profiles imaged by CSST.  Each galaxy profile is characterized by a S\'ersic index ($n$), effective radius ($R_{\rm e}$), position angle (PA), and axis ratio ($b/a$).  These profiles are then used to redistribute the total flux spectrum in two directions and estimate the width of the 2D spectrum in pixels and the self-broadening effect along the dispersion direction due to the lack of slit.  
Second, we use the redistributed 2D spectrum to calculate the number of photon electrons recorded by CSST detectors at the given observing configurations.  We take into account instrumental efficiencies, detector dark currents, and readout noises, as well as background photon noises, to obtain signal and noise spectra in units of electrons.  
Third, we convert these electron numbers into fluxes and combine the signals and noises to generate simulated slitless spectra, as well as signal-to-noise ratio (SNR) spectra for the targets.

Additionally, two modules are included in the workflow.  The first module is used to detect emission lines from the simulated spectra, while the second module is used to estimate the overlap rate and contamination fraction for galaxies in crowded fields.  The latter makes use of the location coordinates, i.e., Right Ascension (R.A.) and Declination (Decl.), of a given list of sources, calculates the coverage of the 2D spectra of these sources within a single CCD detector and estimates the contamination fraction in the group/cluster fields.  By processing the noise from the contamination fluxes of nearby sources, it is possible to quantitatively evaluate detrimental effects on redshift completeness.  More details about these processes are presented in Sections~\ref{sec:methods} and ~\ref{sec:modules}.

\begin{table}
\begin{center}
\resizebox{\columnwidth}{!}{
\begin{tabular}{|l|c|c|c|}
\hline 
Gratings of CSST & $GU$ & $GV$ & $GI$ \\ \hline  \hline

Survey area (deg$^{2}$) &  \multicolumn{3}{c|}{17500}   \\ \hline

Exposures  & \multicolumn{3}{c|}{150\,s\,$\times$\,4} \\ \hline

Point spread function & \multicolumn{3}{c|}{$R_{\rm EE80} \lesssim 0\farcs3$} \\ \hline

Wavelength coverage (nm) & 255\,--\,420 & 400\,--\,650 & 620\,--\,1000 \\ \hline

Spectral resolution [$\lambdaup/(\Delta\lambdaup)$] & 241 & 263 & 270  \\ \hline

5$\sigma$-depth (mag) & 23.2 & 23.4 & 23.2 \\ \hline 

0th to 1st spectrum separation (mm) & 4.22 & 5.60 & 5.44 \\ \hline
\end{tabular}}
\end{center}
\caption[]{The instrumental parameters for the CSST slitless spectroscopic observations.}
\label{tab:parameters}
\end{table}

\subsection{Instrumental parameters}
\label{sec:csstparam}

The slitless spectroscopic observations of CSST will be performed in three bands, namely $GU$, $GV$, and $GI$.  The wavelength coverage and throughput of each CSST slitless grating band are shown in Fig.~\ref{fig:tp_bkg}.  The 5\,$\sigma$ depths for point sources are 23.2, 23.4, and 23.2\,mag in $GU$, $GV$, and $GI$, respectively.  For the CSST wide survey,  a single exposure takes 150\,s and in total four exposures are required at one pointing.  The designed spectral resolution ($R=\lambdaup/\Delta\lambdaup$) of the slitless grating is specified to be in a narrow range of $\gtrsim$200 \citep{Zhan+2021}.  For the CSST slitless spectroscopy, a spectral resolution unit $\Delta \lambdaup$ is designed to cover on average about four pixels on the detector for all three spectroscopic bands.  For simplicity, we adopt a constant spectral resolution for each of the CSST grating as $R=241$ at 337.5\,nm for $GU$, $R=263$ at 525\,nm for $GV$, and $R=270$ at 810\,nm for $GI$ \citep{Zhouxc+2021}.  Therefore, the $\Delta\lambdaup$ above are 1.4, 2.0, and 3.0\,nm for $GU$, $GV$, and $GI$, respectively.  In this way, we have the conversion relation between pixel and wavelength.

CSST gratings generate multiple-order spectra.  The separation distance between the zero-order image and the first-order spectrum is 4.22, 5.60, and 5.44\,mm for $GU$, $GV$, and $GI$, respectively.  The PSF of CSST grating is assumed to be a 2D Gaussian function with a radius of $R_{\rm EE80}$ $\lesssim$ 0\farcs3 enclosing 80\,per\,cent energy on average.  CSST CCD detectors have a pixel size of 10\,$\mu$m and a pixel scale of 0\farcs074, giving the full width at half-maximum (FWHM) of the grating PSF to be four pixels.  The instrumental parameters for the CSST slitless spectroscopic observations are summarized in Table~\ref{tab:parameters}.

\section{Modeling Slitless Spectra}
\label{sec:methods}

We present the methodologies employed in each of the main process steps in \texttt{CESS}.

\subsection{Downgrade Convolution}
\label{sec:convolution}

First, we degrade the high-resolution input galaxy spectra to match the low-resolution CSST spectroscopy.  We re-bin the input spectra to match the CSST grating wavelength resolution.  They are further convolved with a 1D Gaussian kernel with ${\rm FWHM}=\Delta\lambdaup$ to conform with the CSST spectral resolution.  Given the fact that the spectral resolution unit $\Delta\lambdaup$ increases proportionally with the wavelength, we divide each CSST grating band into several wavelength intervals.  The spectra within these intervals are then individually convolved with the corresponding 1D Gaussian kernels.  Classical convolution calculations tend to amplify emission lines when they appear at the boundaries of spectra.  To mitigate this boundary problem, we integrate a weighting function at the borders of neighbouring kernels in our convolution process.  In comparison with accurate pixel-based convolution methods, our refined kernel-based approach maintains the total flux of the convoluted spectra.  Additionally, our method outperforms in terms of computational efficiency.

\subsection{Sky background estimation}
\label{sec:skybkg}

For space-borne optical observations, the sky background is primarily contributed by earthshine and zodiacal light.  When taking slitless spectroscopy, the background photons are also dispersed as the source photons, and the background level may become no longer uniform close to the edges of detectors because a small portion of photons will fall off the detectors.  For CSST, each 9K\,$\times$\,9K CCD detector used for slitless observation is covered by two gratings.  These two gratings are stuck to each other at a certain angle and the connected ridge is higher in front of the detector so that their working orders (first order) disperse towards the centre of the detector \citep{Zhan+2021}.  This design aims to project more spectra within the detector but consequently increase the background level in the central regime covering rough 2K pixel columns in the cross-dispersion direction.

\begin{figure}
\centering
\includegraphics[width=0.48\textwidth]{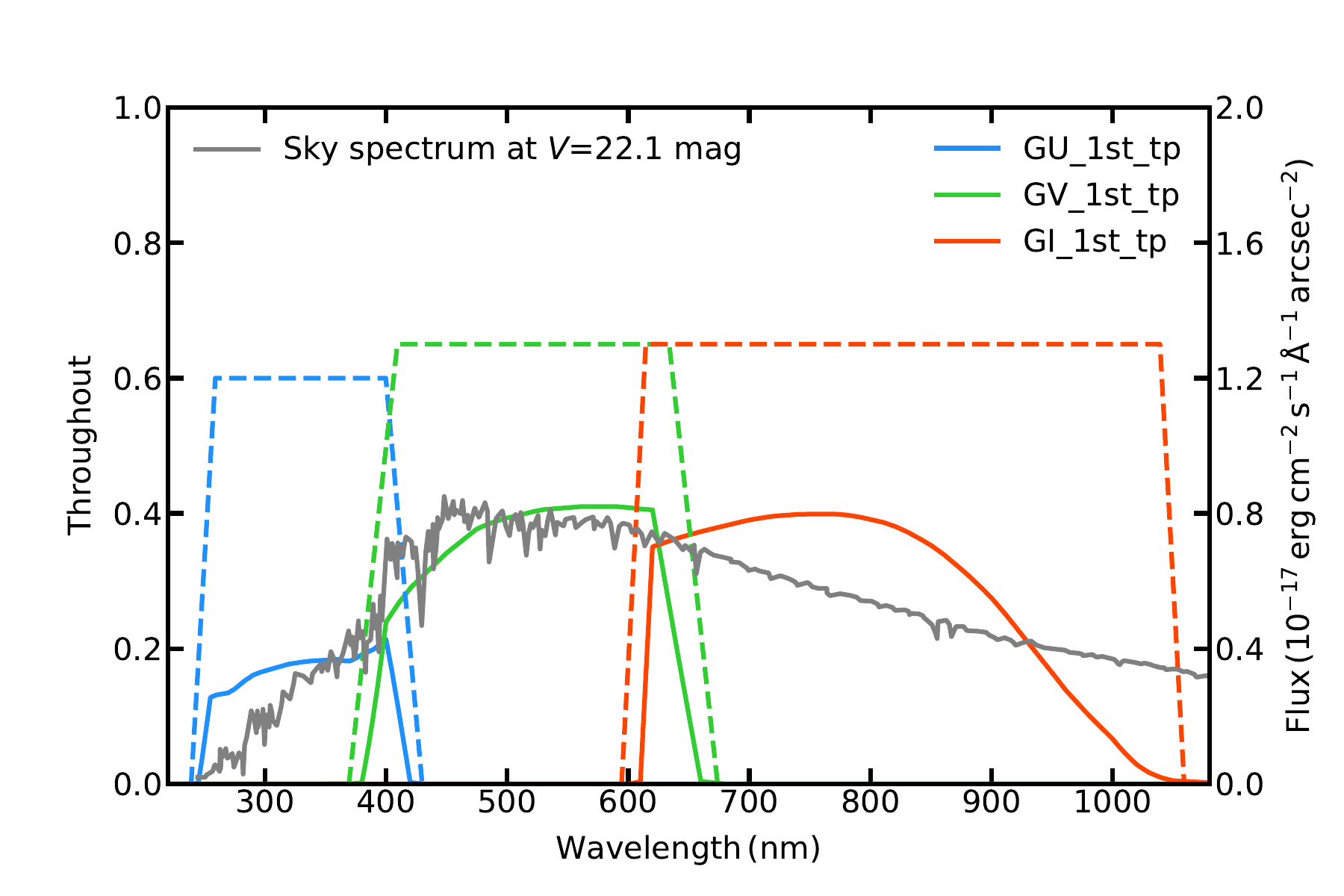}
\caption{ The transmission curves (dashed) and total throughput curves (solid) of CSST slitless gratings.  
The grey solid line stands for the sky spectrum of $V = 22.1$\,mag\,arcsec$^{-2}$ \citep{Dressel+2023}.}
\label{fig:tp_bkg}
\end{figure}

Here we ignore this effect and assume that the sky background is uniformly distributed across the detectors. We estimate the sky background for CSST using sky surface brightness data from HST \citep{Dressel+2023,OBrien+2023}.  In Fig.~\ref{fig:tp_bkg}, we present the sky spectrum at a brightness level of $V=22.1$\,mag\,arcsec$^{-2}$.  The sky background count rate is given by 
\begin{equation} 
\label{eq:bkg}
B_{\rm sky}=A_{\rm eff}\int \tau_{\lambda}\frac{\lambda}{hc} l^2_{\rm p} I_{\rm sky} {\rm d} \lambda,
\end{equation}
where $B_{\rm sky}$ is the sky background count rate in e$^{-}$\,s$^{-1}$\,pix$^{-1}$, $A_{\rm eff}$ is the telescope effective collection area in cm$^2$, $\tau_{\rm \lambdaup}$ refers to the total throughput of a given CSST grating, $h$ is the Planck constant, $c$ is the speed of light, $l_{\rm p}=0\farcs074$ is the pixel scale of the detector, and $I_{\rm sky}$ is the surface brightness of the sky background in units of erg\,s$^{-1}$\,cm$^{-2}$\,\AA$^{-1}$\,arcsec$^{-2}$.  We estimate that $B_{\rm sky}$ is 0.03, 0.36, and 0.50 ${\rm e}^-{\rm s}^{-1}{\rm pixel}^{-1}$ for GU, $GV$, and $GI$ with $I_{\rm sky}(V)=22.1$\,mag\,arcsec$^{-2}$, respectively.  Utilizing the model of zodiacal sky background as a function of heliocentric ecliptic longitude and ecliptic latitude from \citet{Dressel+2023}, we are able to estimate $B_{\rm sky}$ at different sky pointings.

\subsection{Galaxy morphologies}
\label{sec:morphologies}

\begin{figure}
\centering
\includegraphics[width=0.48\textwidth]{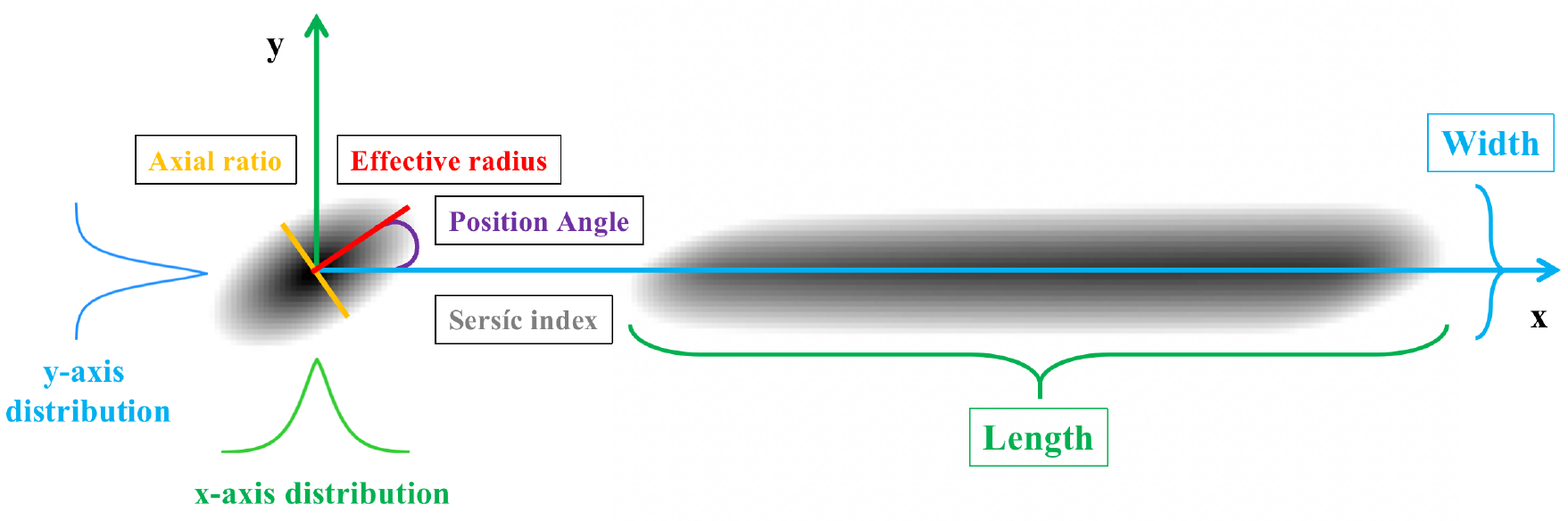}
\caption[123]{An illustration of the resolved 2D galaxy morphology. On the left is the direct image of a galaxy, and on the right is the corresponding 1D slitless spectrum modelled by `\texttt{grizli}'\footnotemark. The 2D morphological parameters including S\'ersic index ($n$), effective radius ($R_{\rm e}$), position angle (PA), and axis ratio ($b/a$) are labelled in the diagram.  The positive PA is defined as the angle from the dispersion direction ($x$-axis) counterclockwise to the major axis of the galaxy.  The $y$-axis (also spatial axis) profile (blue) refers to the projected brightness distribution of the galaxy to the spatial direction, while the $x$-axis profile (green) is the projected brightness distribution to the dispersion direction.  The $y$-axis brightness profile is used to determine the width of the 2D slitless spectrum, and the $x$-axis brightness profile controls the self-broadening effect.}
\label{fig:resovled_2dparam}
\end{figure}

Galaxy images show the 2D brightness profiles that can be quantified by four morphological parameters: S\'ersic index ($n$), effective radius ($R_{\rm e}$), position angle (PA), and axis ratio ($b/a$).  The extended image of a galaxy has prominent effects on its slitless spectrum.  Due to the lack of slits, the light of the entire galaxy image will be dispersed to form a 2D spectrum.  As shown in Fig.~\ref{fig:resovled_2dparam}, the mixture of spatially shifted spectra from different parts of the galaxy image along the dispersion direction induces the self-broadening effect that reduces the spectral resolution.  Similarly, the extended light distribution along a certain PA results in shifts in both the spatial and dispersion directions relative to the centre of the galaxy.

\footnotetext{https://github.com/gbrammer/grizli/}

\begin{table}
\begin{center}
\resizebox{\columnwidth}{!}{

\begin{tabular}{ll}
\hline \hline
Morphological parameters & Sampling values \\ \hline      
S\'ersic index ($n$) & 0.5, 0.6, 0.7, 0.8, 0.9, 1, 1.1, 1.2, 1.3, 1.4, \\
 & 1.5, 1.6, 1.8, 2, 2.5, 3, 3.5, 4, 4.5, 5 \\ \hline 

Effective radius ($R_{\rm e}$) (arcsec) & 0.3, 0.5, 0.7, 0.9, 1, 1.2, 1.4, 1.6, 1.8, 2, \\
 & 2.5, 3, 3.5, 4.5, 5, 5.5, 6, 6.5, 7, 7.4 \\ \hline

Position angle (PA) ($^{\circ}$) & 0, 10, 20, 30, 40, 50, 60, 70, 80, 90 \\ \hline

Axis ratio (b/a) & 0.1, 0.2, 0.3, 0.4, 0.5, 0.6, 0.7, 0.8, 0.9, 1 \\ \hline
\end{tabular}}
\end{center}
\caption[]{Settings of morphological parameters used in the pre-established 2D profile library.}
\label{tab:morphological_parameters}
\end{table}

From the 2D brightness profile of a galaxy, we derive two brightness distribution functions: one along the dispersion axis and the other along the vertical axis, representing the spatial direction.  The former serves as a weight function and also a wavelength offset function, enabling the co-addition of the same intrinsic slitless spectra to quantify the self-broadening effect.  Meanwhile, the vertical brightness distribution function determines the width of the extraction window for obtaining the 1D spectrum from the 2D spectrum image.  A criterion of ${\rm SNR}>1$ is adopted to identify signal pixels and decide the averaged width and length of the 2D spectrum image.

It's worth noting that, in principle, correction for the $x$-axis shift of the major axis is doable when summing up the vertical pixel columns of the 2D spectrum.  Our tests indicate that this correction significantly mitigates the broadening effect, particularly for extended galaxies (with $R_{\rm e}\gg$ PSF) with small axis ratios (close to edge-on).  These extended galaxies are often bright.  In practice, focusing on the central part of extended sources helps diminish the broadening effect.  In our input catalogue, the vast majority of galaxies exhibit $R_{\rm e}<\sim$PSF (0\farcs 3) (see Section~\ref{sec:applications}).  Consequently, the additional broadening effect caused by the misalignment of the major axis of galaxies is negligible.

Additionally, galaxy kinematics contribute to spectral shifts in the dispersion direction. However, considering that the velocities of galaxies are substantially smaller than the resolved minimal velocity at $R\gtrsim 200$ of the CSST grating spectroscopy, we disregard the effect of galaxy kinematics on the 1D spectra.

Many photometric and spectroscopic surveys are carried out with ground-based facilities and lack accurate morphological measurements for distant galaxies.  We utilize a catalogue of galaxies with known morphological parameters to establish empirical relations between the galaxy morphology and other galaxy parameters such as magnitudes, redshift, and stellar mass.  Doing so we are able to simulate galaxy images for the input galaxies with structural parameters in \texttt{CESS}.

We generate a library of galaxy images matching CSST grating PSF (FWHM=0\farcs3).  The library contains galaxies with representative morphology parameters:  20 $n$ in the range of [0.5, 5], 20 $R_{\rm e}$ in the range of [0\farcs3, 7\farcs4], 10 PA in the range of [0$^{\circ}$, 90$^{\circ}$] with an interval of 10$^{\circ}$, and 10 $b/a$ in the range of [0, 1] with an interval of 0.1. These parameters of the library are summarized in Table~\ref{tab:morphological_parameters}.  In total, we have 40,000 simulated 2D brightness profiles of galaxies at a spatial resolution of PSF FWHM=0\farcs 3.

The GEMS morphological catalogue from \citet{Haussler+2007} and the mass-size relations from \citet{Shen+2003} and \citet{vanderWel+2014} are adopted to match each of the input galaxies with a galaxy image from the library of the 2D brightness profiles of different $n$, $R_{\rm e}$, PA and $b/a$.  The GEMS catalogue contains more than 40,000 galaxies with $R<24$\,mag over $0<z<1$ with morphological parameters measured from \textit{HST} F850LP imaging.  We divide GEMS galaxies into subsamples by magnitude, PA, $b/a$, and spectral type (quiescent versus star-forming classified by $n$).  Then we derive the corresponding probability distributions of the subsamples to pick up $n$, PA, and $b/a$ for the input galaxies.  Utilizing the known redshift, stellar mass, and spectral type of each input galaxy, we infer the $R_{\rm e}$ with the mass-size relations.  With the four morphological parameters $n$, $R_{\rm e}$, PA, and $b/a$ of each input galaxy, we go to our library of galaxy images and take the best-matching one as the model galaxy image for further analysis.

\begin{figure*}
\centering
\includegraphics[width=0.98\textwidth]{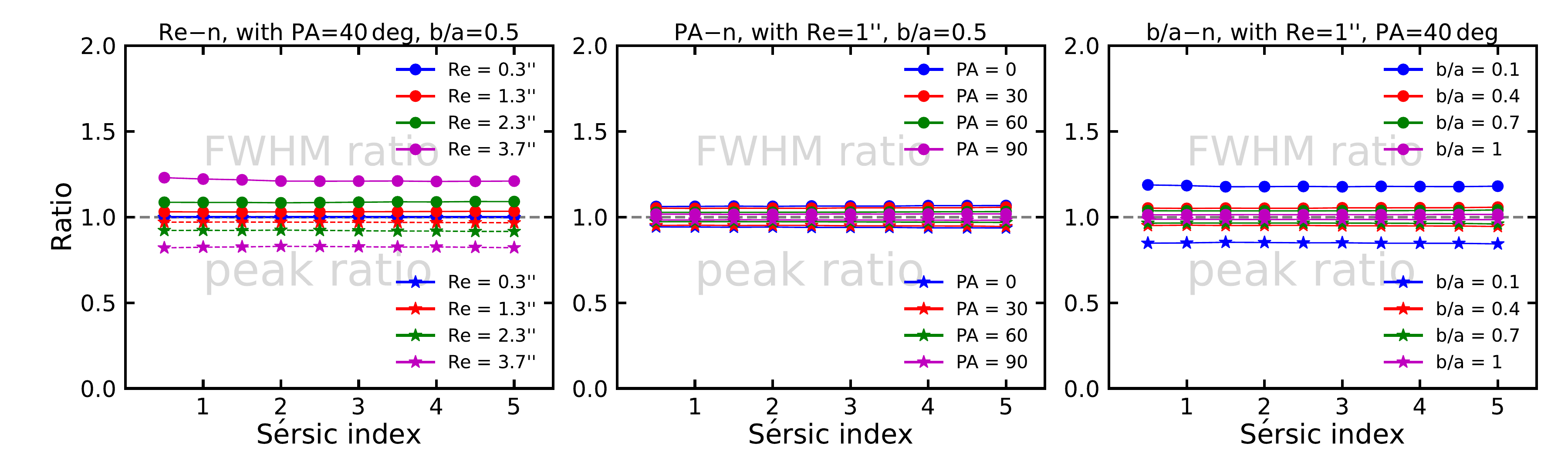}
\includegraphics[width=0.98\textwidth]{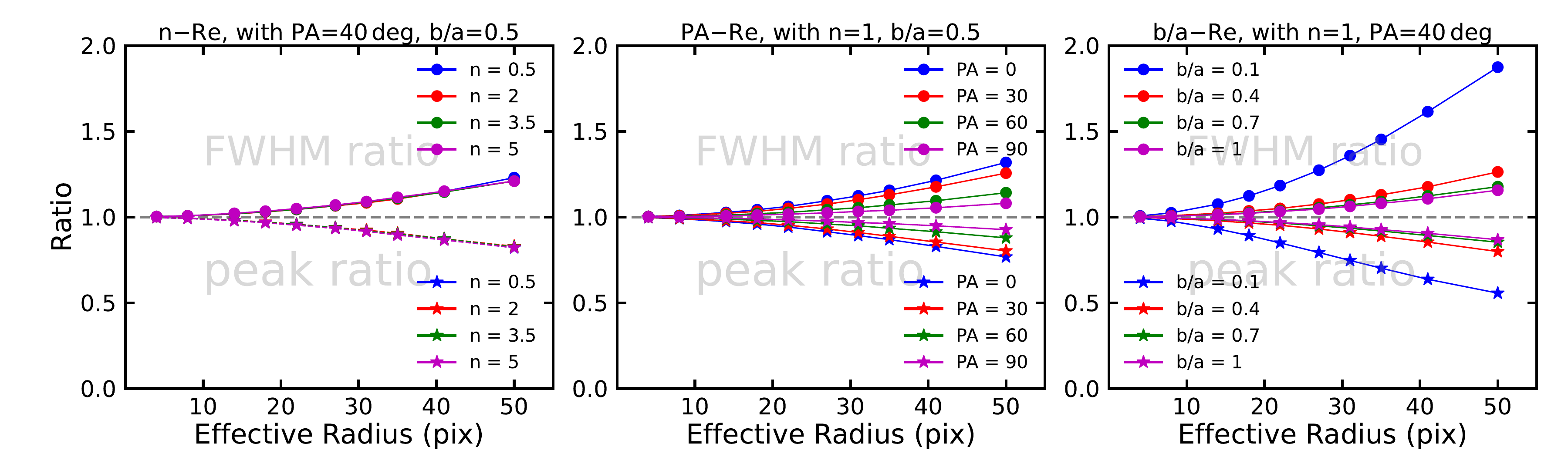}
\includegraphics[width=0.98\textwidth]{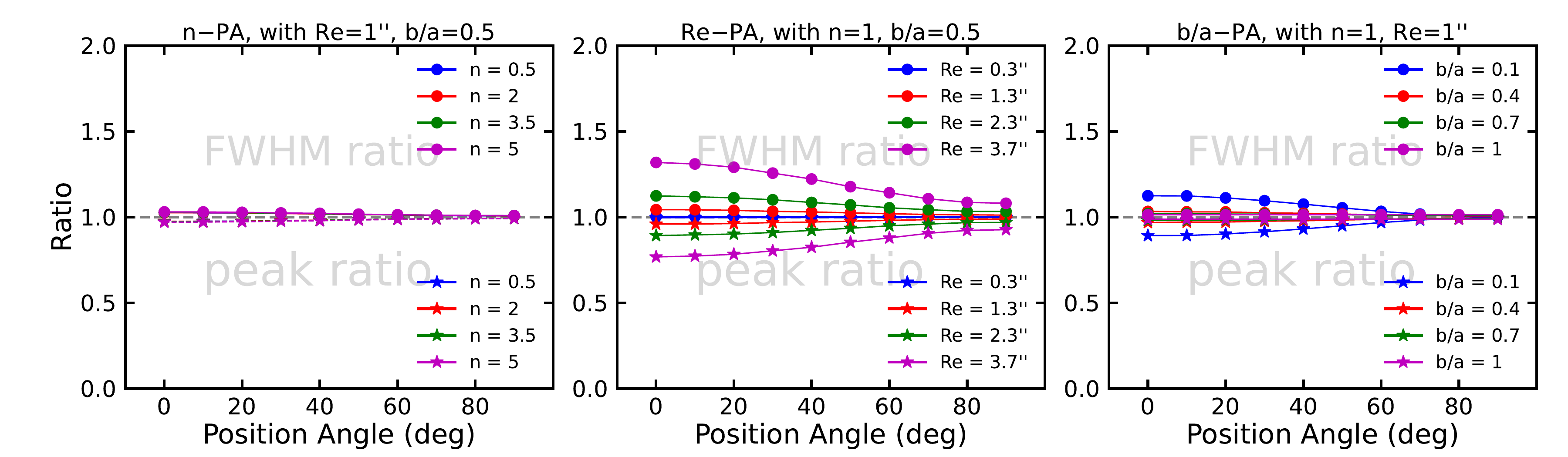}
\includegraphics[width=0.98\textwidth]{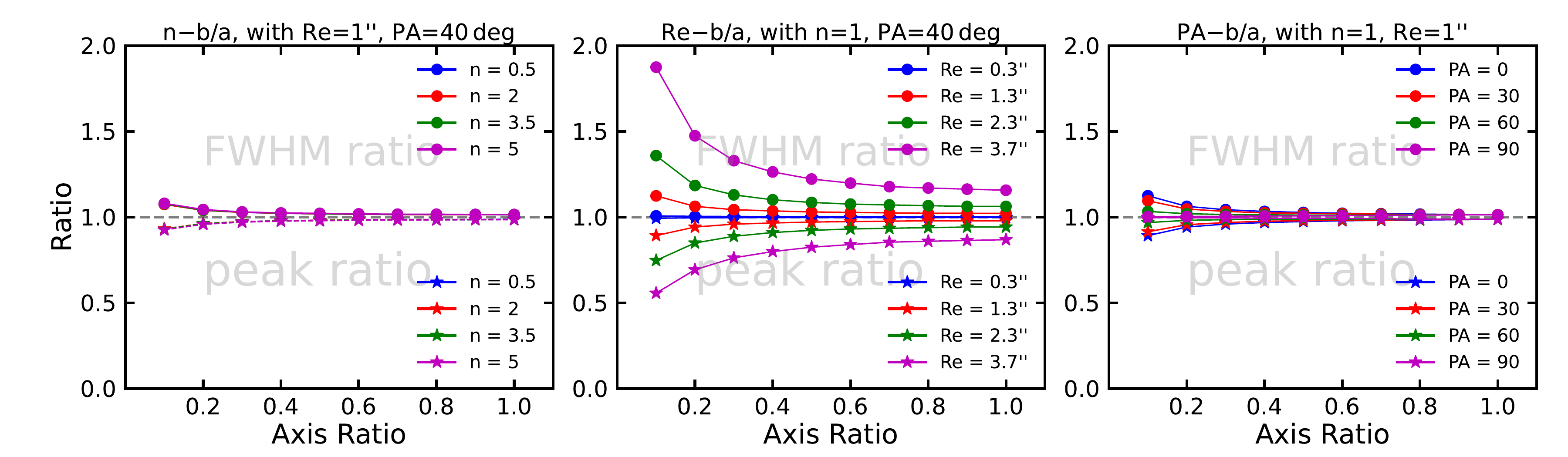}
\caption{Demonstration of the morphological broadening effects on emission lines of slitless spectra.  The broadening leads to an increase in FWHM and a decrease in the peak intensity for emission lines.  The ratio of the broadened to the original value is used to quantify the change in FWHM (circles) or peak intensity (stars).  From top to bottom panels, the two ratios are shown as a function of S\'ersic index ($n$), effective radius ($R_{\rm e}$), position angle (PA), and axis ratio ($b/a$), respectively.  In each of the three panels of the same parameter in the $x$-axis,  the dependent relations on the second morphological parameters are also presented while the other two parameters are fixed.  At $R_{\rm e} \gtrsim 1$\,arcsec (15\,pix), the broadening effects start to be increasing.  Additional broadening effects could be added when galaxies are viewed at nearly edge-on ($b/a<0.3$).  One can see that $n$ has little or no contribution to the morphological broadening effect.  It is clear that the broadening effect is primarily dominated by $R_{\rm e}$, while PA and $b/a$ play less important roles.} 
\label{fig:el_evolution}
\end{figure*}

In short, we obtain the four morphological parameters for all galaxies in the input catalogue using the empirical relations derived from the GEMS catalogue and mass-size relations.  Using $n$, $R_{\rm e}$, PA, and $b/a$, we assign each of the target galaxies with the corresponding 2D brightness profile through a nearest neighbour search in the pre-established library of galaxy images.  Such an approach is also aimed at saving computation time and achieving the goal of enabling our program \texttt{CESS} to quickly generate a large number of simulated CSST slitless spectra.

\subsection{The Self-broadening effect for extended sources}
\label{sec:broadening}

All beams of light will be dispersed when passing through a slitless grating.  The dispersed spectra from the closely connected beams along the dispersion axis will mix together.  This would form a blending spectrum and leave the spectral features (e.g. emission lines) broader than those at the instrumental spectral resolution.  This self-broadening effect in the dispersion direction is driven by the galaxy morphology and can be quantified through simulations.

For simplicity, we conduct an artificial spectrum template and add emission lines using the standard Gaussian function and matching the spectral resolution of $R=200$.  To achieve the simulation, our approach involves obtaining the projected brightness profile along the dispersion direction of the object, which is then utilized as a weight function to quantify the self-blending effect of the slitless spectrum (see Fig.~\ref{fig:resovled_2dparam}).  To better illustrate the impact of morphological broadening, we focus on the changes in FWHM and peak intensity of emission lines between the broadened and original states.  We generate mock spectra as functions of galaxy morphological parameters and investigate the changes in the FWHM ratio and the peak intensity ratio.  Fig.~\ref{fig:el_evolution} presents the results of our simulations on the self-broadening effects.  Note that the total flux of a line is conserved in our simulations and the increase in FWHM is always tied with a decrease in peak intensity.  We therefore focus on the changes in the FWHM ratio for clarification.

Our simulation results show that at increasing $n$ (the first row of Fig.~\ref{fig:el_evolution}), the FWHM ratio remains constant, and this holds with different $R_{\rm e}$, PA, and $b/a$.  Meanwhile, the FWHM ratio apparently increases at increasing $R_{\rm e}$ and at decreasing $b/a$, and marginally increases at decreasing PA.  These suggest that $n$ has minimal impact on the broadening of the emission lines.  From a broad perspective, these trends are regulated by $R_{\rm e}$, $b/a$, and PA, irrelevant with $n$.  In order to study the evolutionary relationship between each pair among these four parameters, typical values for the parameters are used when they should be constant.

Delving deeper into the trends of the other three parameters, for example, the FWHM ratio rapidly increases at increasing effective radius ($R_{\rm e}$) (the second row of Fig.~\ref{fig:el_evolution}).  The broadening effect starts to become significant at $R_{\rm e} \gtrsim 1$\,arcsec (15\,pix).  Keeping other parameters relatively moderate, when the PSF increases to three times the original value, the ratio of FWHM increases by approximately 2\,per\,cent.  When the PSF continues to grow, the ratio of FWHM can increase by about 10--30\,per\,cent.  With a constant $R_{\rm e}$, the increase of the FWHM ratio becomes larger at decreasing $b/a$ (from face-on to edge-on) as well as decreasing PA (from perpendicular to parallel for the major axis in relation to the dispersion axis).  For an extreme case, the influence of $R_{\rm e}$ on the variation of FWHM can reach more than 80\,per\,cent when $b/a$ approaches 0.1.

\begin{figure*}
\centering
\includegraphics[width=0.45\textwidth]{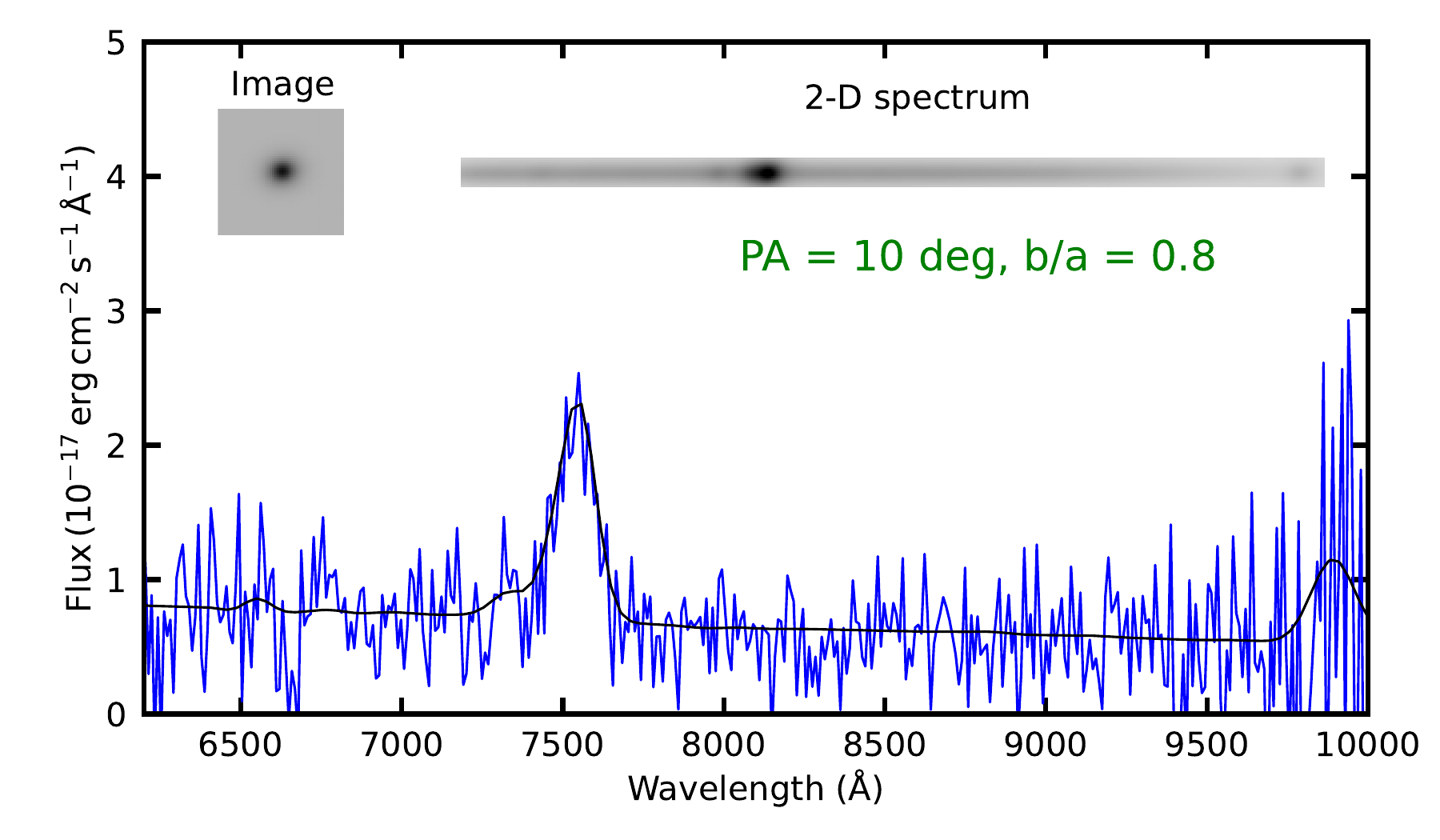}
\includegraphics[width=0.45\textwidth]{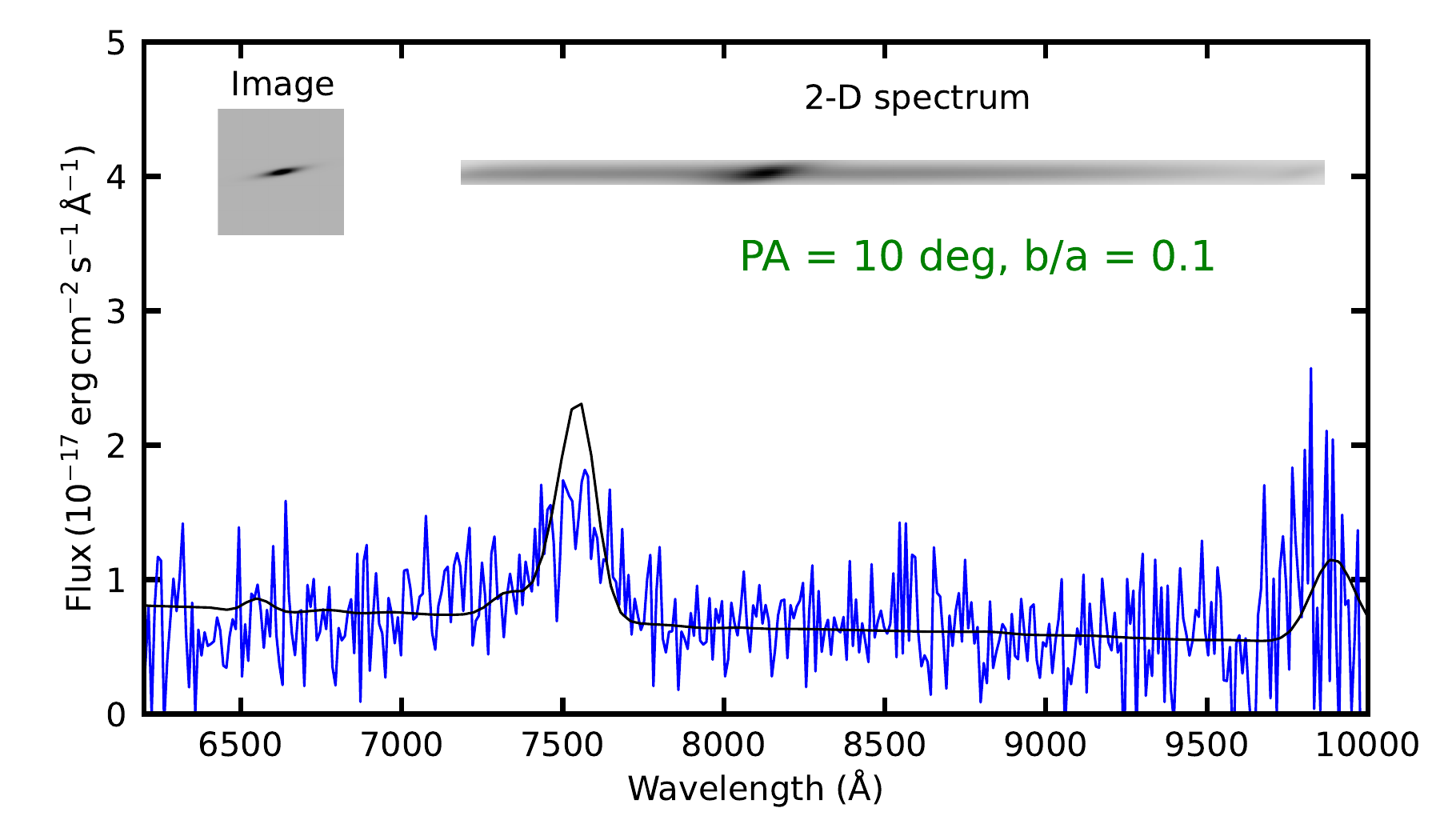}
\includegraphics[width=0.45\textwidth]{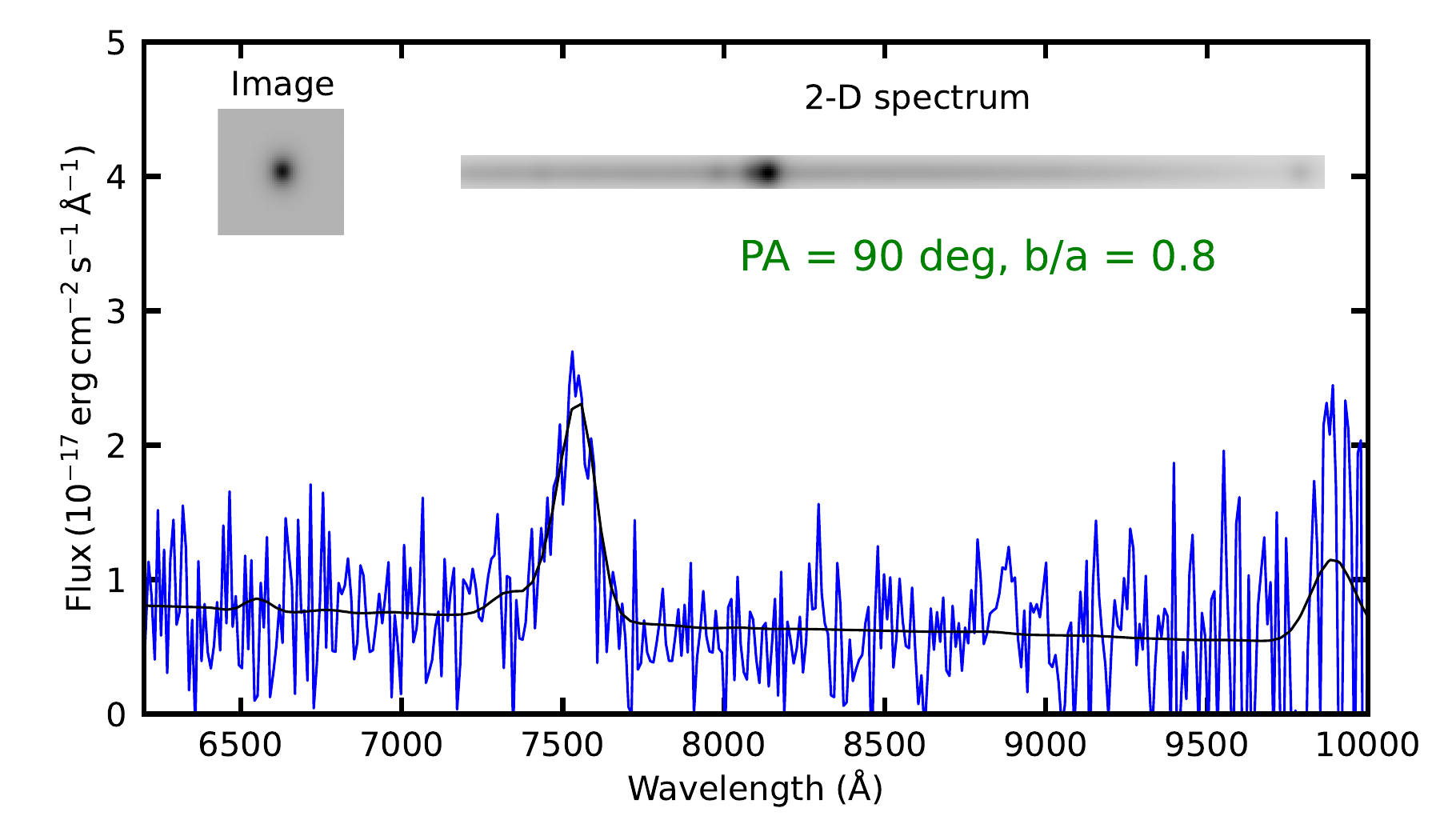}
\includegraphics[width=0.45\textwidth]{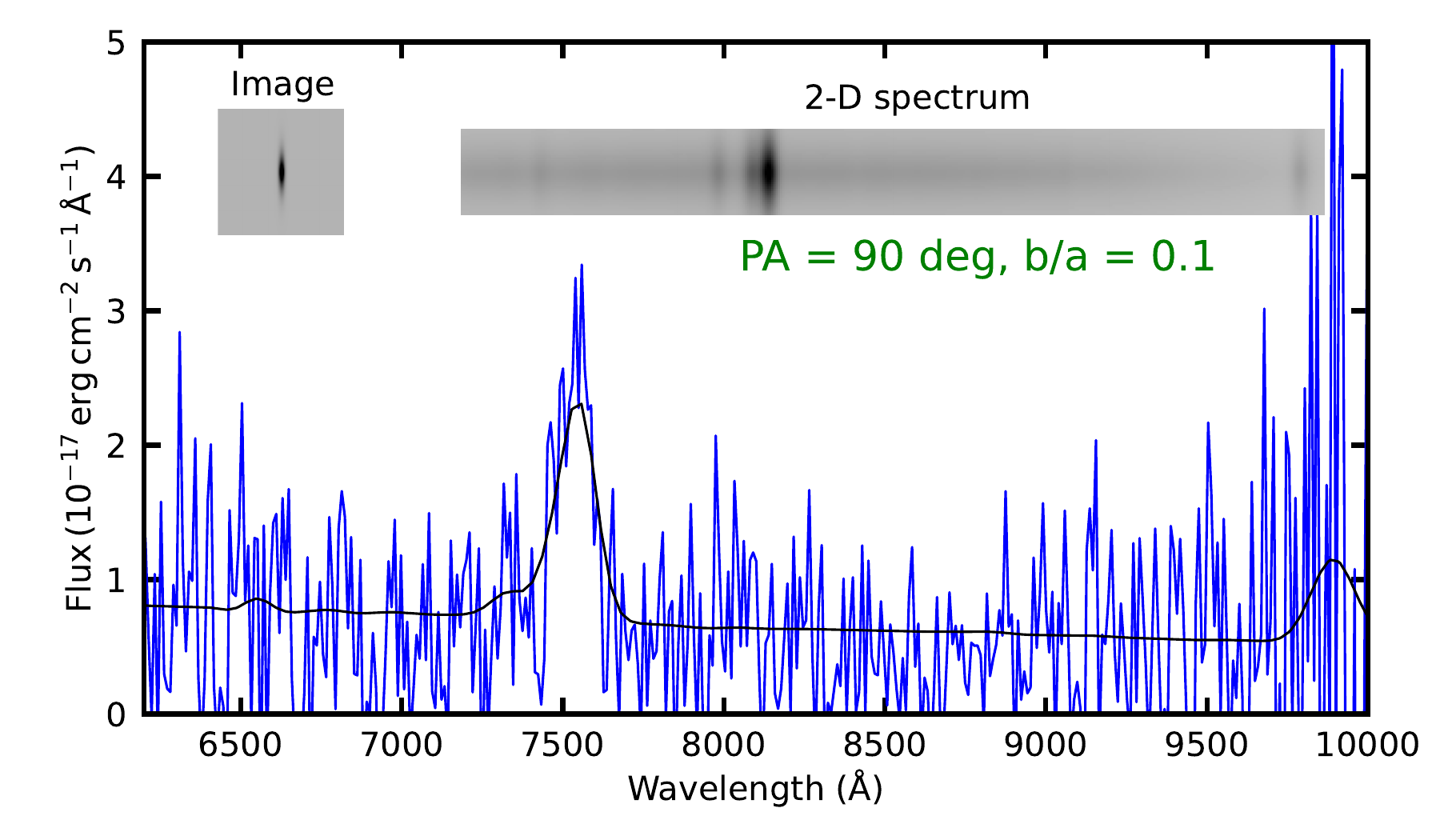}
\caption{Examples of simulated 1D and 2D slitless spectra and images for the same galaxy ($R_{\rm e}=5$\,pix and $n=1$) viewed at four different angles.  The comparison demonstrates how galaxy morphologies influence the peak intensity and width of emission lines.  The simulated 1D slitless spectra have a strong \OIII\ emission line at $z=0.51$.  The black solid lines represent the same simulated signal spectrum of the original galaxy, while the blue solid lines refer to the simulated spectra with noise and the morphological broadening effect taken into account.  
In each panel, the 2D galaxy image and the 2D slitless spectrum image (both without noise) are shown.  The width of the 2D spectrum image varies with the 2D morphology.  }
\label{fig:2d_spec_example}
\end{figure*}

Position angle (PA) (the third row of Fig.~\ref{fig:el_evolution}) starts to play a role when either the effective radius is large or the axis ratio is small (nearly edge-on).  When the major axis is nearly parallel to the dispersion axis, the dispersion direction convolves more flux, intensifying the broadening effect of the slitless spectrum.  The FWHM ratio increases up to about 30\,per\,cent with $R_{\rm e} = 50\,$pix and 10\,per\,cent with $b/a = 0.1$ when PA approaches 0$^{\circ}$.  As for axis ratio ($b/a$) (the fourth row of Fig.~\ref{fig:el_evolution}), the broadening effect begins to emerge at $b/a<0.3$ and rises to its max when the source is close to edge-on.  It slightly affects the FWHM ratio to increase up to about 8--13\,per\,cent with given $n$ and PA, but significantly affects the ratio when $R_{\rm e}$ is large.  The ratio quickly grows to more than 1.8 at $b/a=0.1$ when $R_{\rm e}=50$\,pix.

In summary, among the four morphological parameters,  S\'ersic index has little or no contribution, and effective radius is the dominant factor to the morphological broadening effect, while axis ratio and position angle play less important roles.  Moreover, when $R_{\rm e}$ is large, the impacts of small axis ratios (nearly edge-on) or small position angles (nearly parallel to the dispersion axis) on the broadening may become significant.

\subsection{Noise estimation} 
\label{sec:noise}

Using the intrinsic flux-calibrated slitless 1D spectrum and the 2D brightness profile of a galaxy, we can calculate the count rate of electrons recorded by CCD detectors following 
\begin{equation}
\label{eq:countrate}
C=A_{\rm eff}\int_{\lambda_{\rm min}}^{\lambda_{\rm max}} \tau_{\lambda}\frac{\lambda}{hc} F_{\lambda}{\rm d} \lambda,
\end{equation}
where $C$ is the electron count rate of the source in e$^-$\,s$^{-1}$ and $F_{\lambdaup}$ represents galaxy flux in units of erg\,s$^{-1}$\,cm$^{-2}$\,\AA$^{-1}$.  This formula can estimate the electron count rates in a spectral resolution unit $\Delta \lambdaup$, and then obtain the electron counts of each resolution unit by multiplying the total exposure time (150\,s$\times$4$\,=\,$600\,s).  We also introduce a typical error of 0.1\,per\,cent in wavelength calibration, which is about 0.4\,pix ($\sim0.1$\,$R_{\rm EE80}$).

Here we derive signal from the electron counts from the source as $C \times t$, calculate the root mean square (RMS) of photon noises from source and background, dark current, and readout noises, and yield an SNR in a given spectral resolution unit as 
\begin{equation}
\label{eq:SNR}
{\rm SNR}=\frac{C\times t}{\sqrt{C \times t + N_{\rm pix} \times (B_{\rm sky} + B_{\rm dark}) \times t + N_{\rm pix} \times N_{\rm read} \times R_{\rm n}^2}},
\end{equation}
where $N_{\rm pix}$ is number of pixels within the extraction box of the 2D spectrum in a given spectral resolution unit, $B_{\rm dark}=0.02$\,e$^-$\,s$^{-1}$\,pix$^{-1}$ is the detector dark current, $N_{\rm read}=4$ is number of exposures, and $R_{\rm n}=5$\,e$^-$\,pix$^{-1}$ is the readout noise.   The SNR calculated here is defined as all signal overall noise in the given spectral resolution unit, which is typical 4\,pix (spectral, $\sim$PSF) $\times$ the width of the 2D slitless spectrum (spatial).  Given the CSST instrumental parameters (see Section~\ref{sec:csstparam}), we estimate the number of spectral resolution units of each grating band, and the result is 127, 142, and 148 for $GU$, $GV$, and $GI$, respectively.

Doing so we obtain spectra of signal, noise RMS, and SNR in units of electron count.  We take the noise RMS to randomly generate noise satisfying the normal distribution and add the noise to the signal spectrum.  By Converting the spectrum of signal\,+\,noise from electron count into flux, we obtain the simulated CSST 1D slitless spectrum for the target galaxy, together with its SNR spectrum, and the noise RMS spectrum.

One emission-line galaxy (ELG) is used to provide a more intuitive representation of the impact of the self-blending effects of PA and $b/a$ and the results are shown in Fig.~\ref{fig:2d_spec_example}.  The original galaxy has a strong \OIII $\lambdaup\lambdaup$4959, 5007 emission line in $GI$ with a redshift at 0.51 and a morphology property of $n = 1$, $R_{\rm e} = 5$\,pix.  The sample galaxy is viewed at nearly face-on (${\rm PA} = 10^{\circ}$ and $b/a = 0.8$) in the top left panel, at edge-on and nearly parallel to the dispersion axis (${\rm PA} =10^{\circ}$ and $b/a=0.1$) in the top right panel, at edge-on and perpendicular to the dispersion axis (${\rm PA} =90^{\circ}$ and $b/a=0.1$) in the bottom right panel, and at nearly face-on with ${\rm PA} = 90^{\circ}$ and $b/a=0.8$ in the bottom left panel.  
Note that in our program, we generate the 2D slitless spectra using the python package `\texttt{sls\_1d\_spec}'\footnote{https://csst-tb.bao.ac.cn/code/zhangxin/sls\_1d\_spec/}.  
The results demonstrated in Fig.~\ref{fig:2d_spec_example} confirm our conclusions on how the morphological parameters decide the broadening effect.  Again, galaxy morphologies viewed at edge-on and with the major axis parallel to the dispersion axis will maximize the morphological broadening effect. In contrast, observing disc galaxies at edge-on with the major axis perpendicular to the dispersion axis will minimize the morphological broadening effect.

\section{Ancillary Modules}
\label{sec:modules}

We develop two additional modules to analyse the simulated slitless 1D spectra for our scientific goals.  One module is aimed at detecting emission lines to test redshift measurements for star-forming galaxies.  The other is used to evaluate the failure rate of redshift identification for galaxies in crowded fields like clusters of galaxies.

\subsection{Detection for emission lines}
\label{sec:EL_detection}

\begin{figure}
\centering
\includegraphics[width=\linewidth, angle=0]{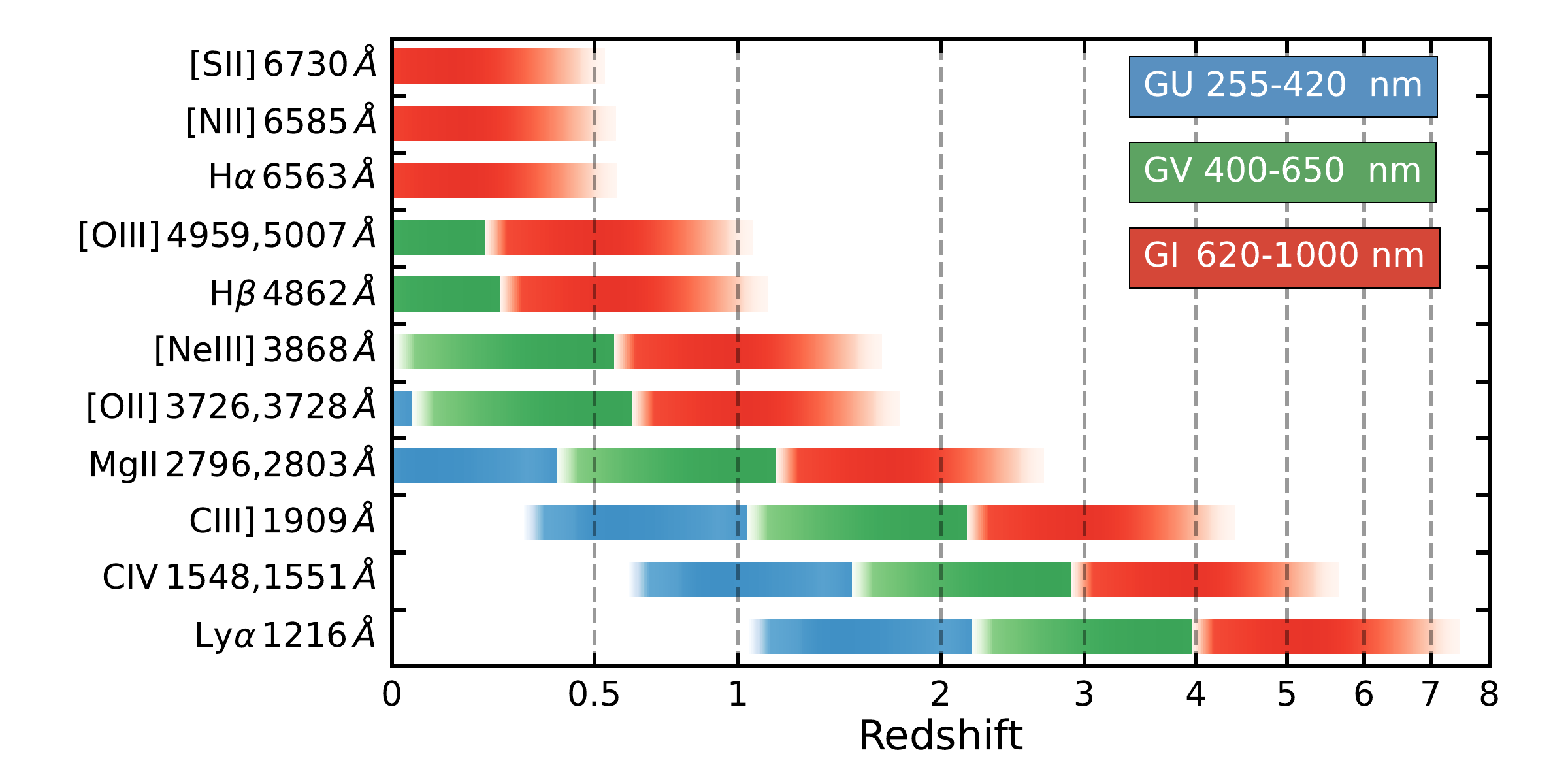}
\caption{ The redshift ranges for strong emission lines detected in CSST slitless grating $GU$ (blue), $GV$ (green), and $GI$ (red), respectively.  The wavelength coverages of the three bands are marked.}
\label{fig:zcoverage}
\end{figure}

For galaxy redshift surveys, emission lines are crucial spectral features to reliably measure redshifts.  Fig.~\ref{fig:zcoverage} illustrates the redshift range of major emission lines in the UV to optical bands covered by three CSST gratings.  The CSST slitless spectroscopic observations allow us to probe optical emission lines of galaxies at $z<1.5$ and identify high-$z$ galaxies with Ly$\alpha$ emission line.

ELGs are the most abundant galaxies in the Universe and act as key tracers of the cosmic large-scale structures for measuring cosmological parameters \citep{Jimenez+2021}.  For instance, \citet{Gong+2020} used the line intensity power spectrum to constrain the cosmological parameters; \citet{Ivanov+2021} made use of the full-shape analysis of power spectra based on effective field theory for ELGs to measure cosmological parameters.  The redshift measurements of ELGs are more accurate because of the presence of emission lines.

\begin{figure}
\centering
\includegraphics[width=1\linewidth, angle=0]{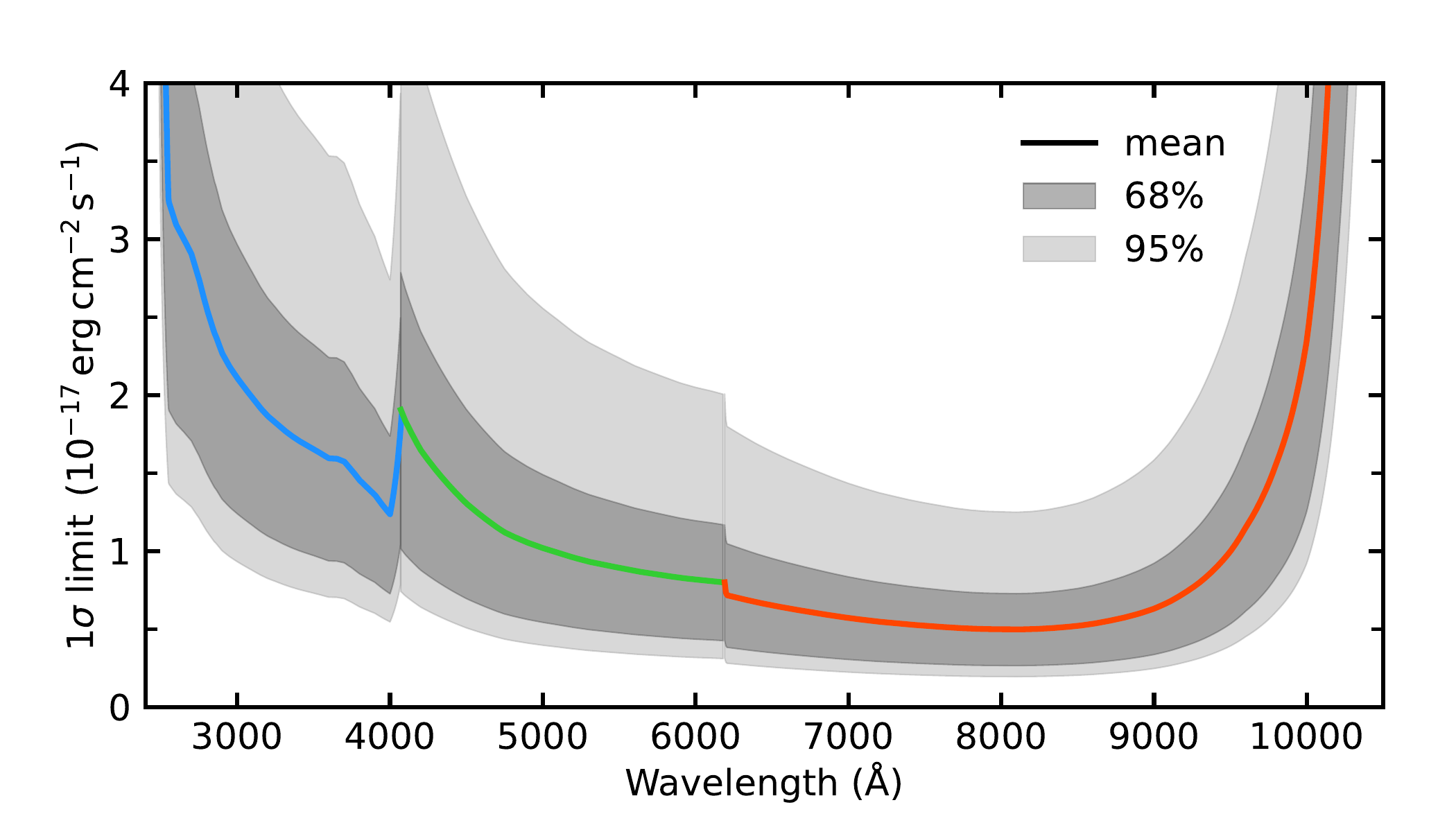}
\caption{ The 1$\sigma$ sensitivity curves for the three CSST slitless gratings. The dark grey and light grey shadows show the 68 and 95\,per\,cent spread of the limits for the 100,000 sample spectra.  The thick solid line shows the mean 1$\sigma$ sensitivities of  $GU$ (blue), $GV$ (green), and $GI$ (red), respectively.}
\label{fig:1sigmalimit}
\end{figure}

By estimating the 1$\sigma$ flux limits stepping through $GU$, $GV$, and $GI$, we estimate the line flux sensitivity of CSST slitless spectra as shown for 100,000 spectra in Fig.~\ref{fig:1sigmalimit}.  For compact sources, we used extraction apertures of 8 (spatial) by 4 (spectral) pixels to derive the 1$\sigma$ limit for each band.  The mean 1$\sigma$ noise level reaches about 1.5, 1, and 0.5$\times$10$^{-17}$\,erg\,s$^{-1}$\,cm$^{-2}$ for $GU$, $GV$, and $GI$, respectively.   For more extended emission and sources, a larger aperture of, for example, 16 (spatial) by 8 (spectral) pixels, might be more representative of the sensitivity estimates.  Such an aperture will decrease the sensitivity by a factor of 2.

Program $\texttt{find\_lines\_derivative}$ in the python package \texttt{specutils}\footnote{https://specutils.readthedocs.io/en/stable/} is adopted by our program \texttt{CESS} to detect the emission lines in the simulated slitless 1D spectra.  This line-detection program adopts functions of derivatives and thresholding fluxes to detect emission or absorption lines in a spectrum.  Doing so, \texttt{CESS} is able to detect the presence of emission lines in the simulated slitless 1D spectra.  We apply the detection for emission lines to both the intrinsic spectra (pure signal without noises) and the simulated spectra (signal\,+\,noise).  The detections in the intrinsic spectra are considered as genuine ones. The emission lines detected in the simulated signal+noise spectra are recognized as true detections once they are also in the vicinity of the emission lines detected in the intrinsic spectra.  Based on the signal and noise spectra, we are able to calculate SNR for the detected emission lines and measure the strength and reliability of the detected emission lines in the simulated spectra.  The line detection results, including the central wavelengths, fluxes, and SNR are written into an output line catalogue.

\begin{figure}
\centering
\includegraphics[width=0.48\textwidth, angle=0]{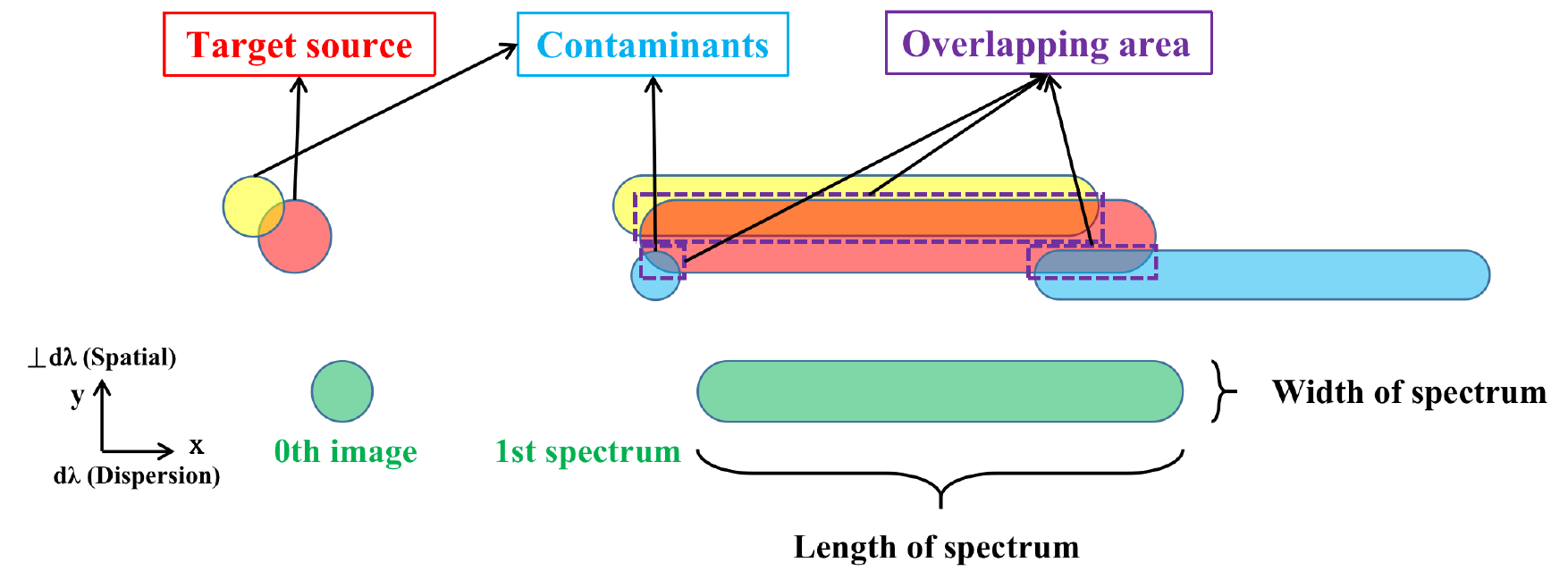}
\caption{ An illustration of overlap contamination in the slitless spectroscopy.  The relatively isolated source (green) shows the case without contamination from the overlapping effect.  The extraction areas of the 2D spectra can be determined by \texttt{CESS} following the instrumental parameters listed in Table~\ref{tab:parameters}.  The target source (red) is contaminated by two adjacent sources shown in yellow and blue, respectively.  The fluxes of adjacent sources within the extraction area of the target spectrum, including the zero-order images and the first-order spectra, are taken as contamination fluxes to estimate additional noises. }
\label{fig:overlap}
\end{figure}

\begin{figure*}
\centering
    \begin{subfigure}[b]{0.48\textwidth}
        \centering
        \includegraphics[width=\linewidth, angle=0]{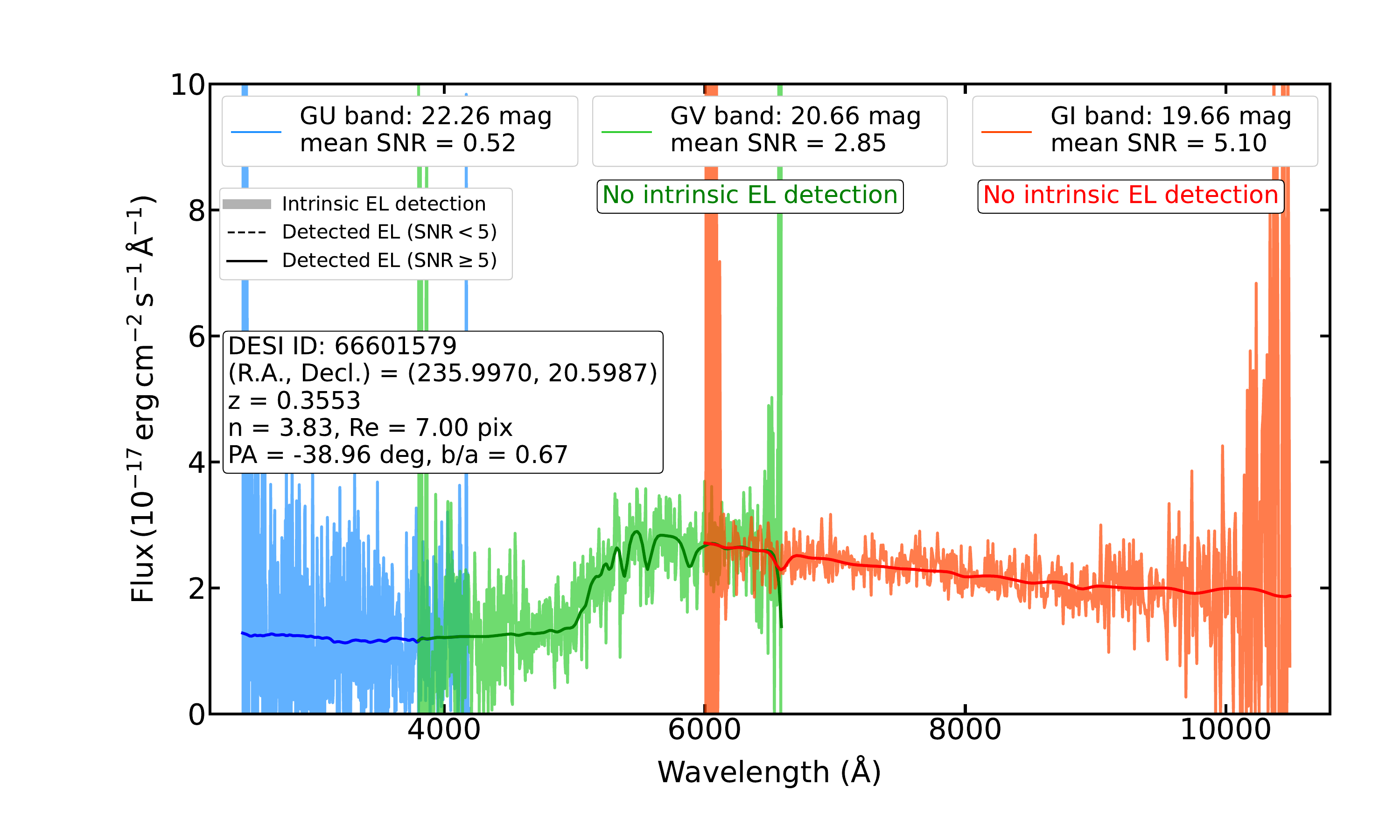}
        \caption{No Emission Line}
    \end{subfigure}
    \hfill
    \begin{subfigure}[b]{0.48\textwidth}
        \centering
        \includegraphics[width=\linewidth, angle=0]{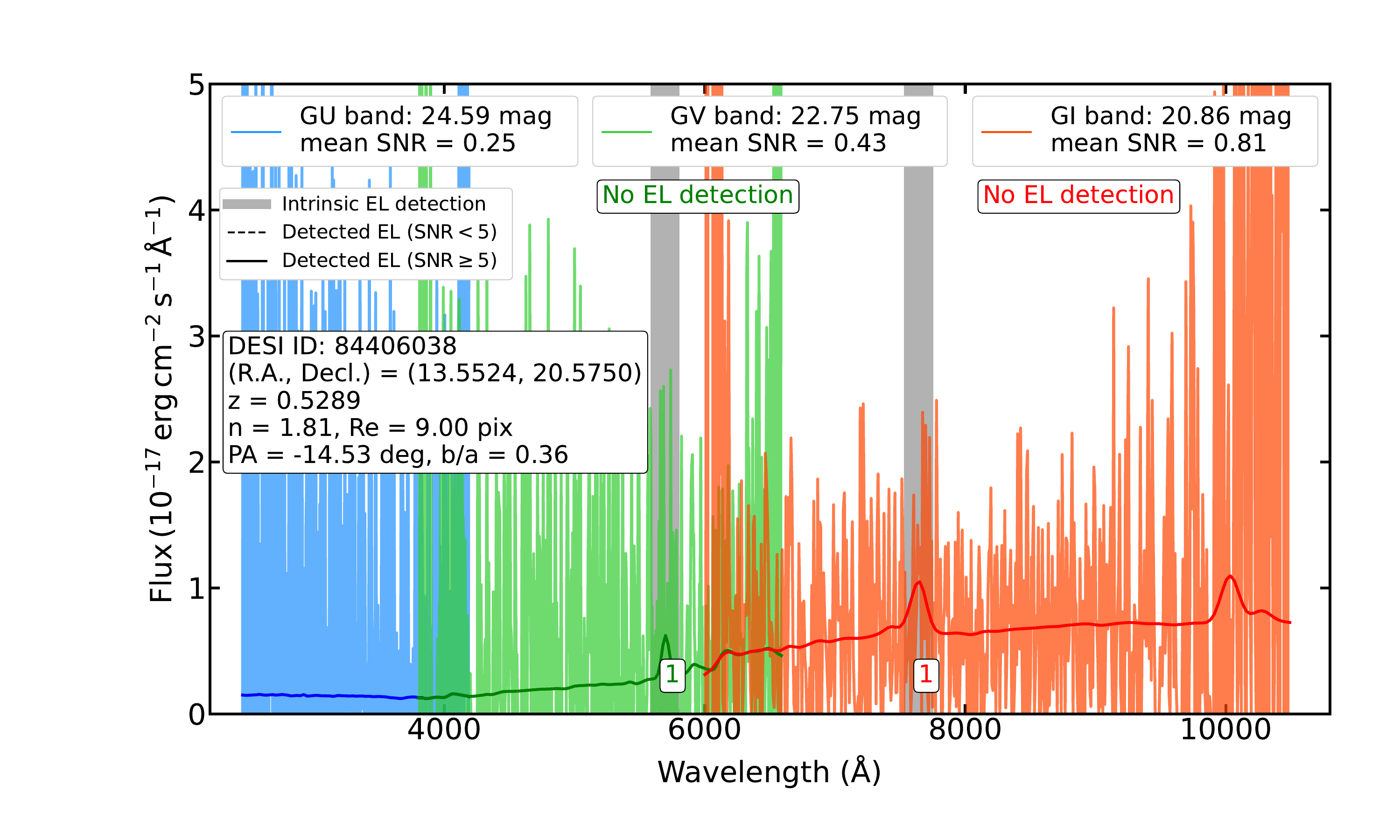}
        \caption{No EL detection}
    \end{subfigure}
    \hfill
    \begin{subfigure}[b]{0.48\textwidth}
        \centering
        \includegraphics[width=\linewidth, angle=0]{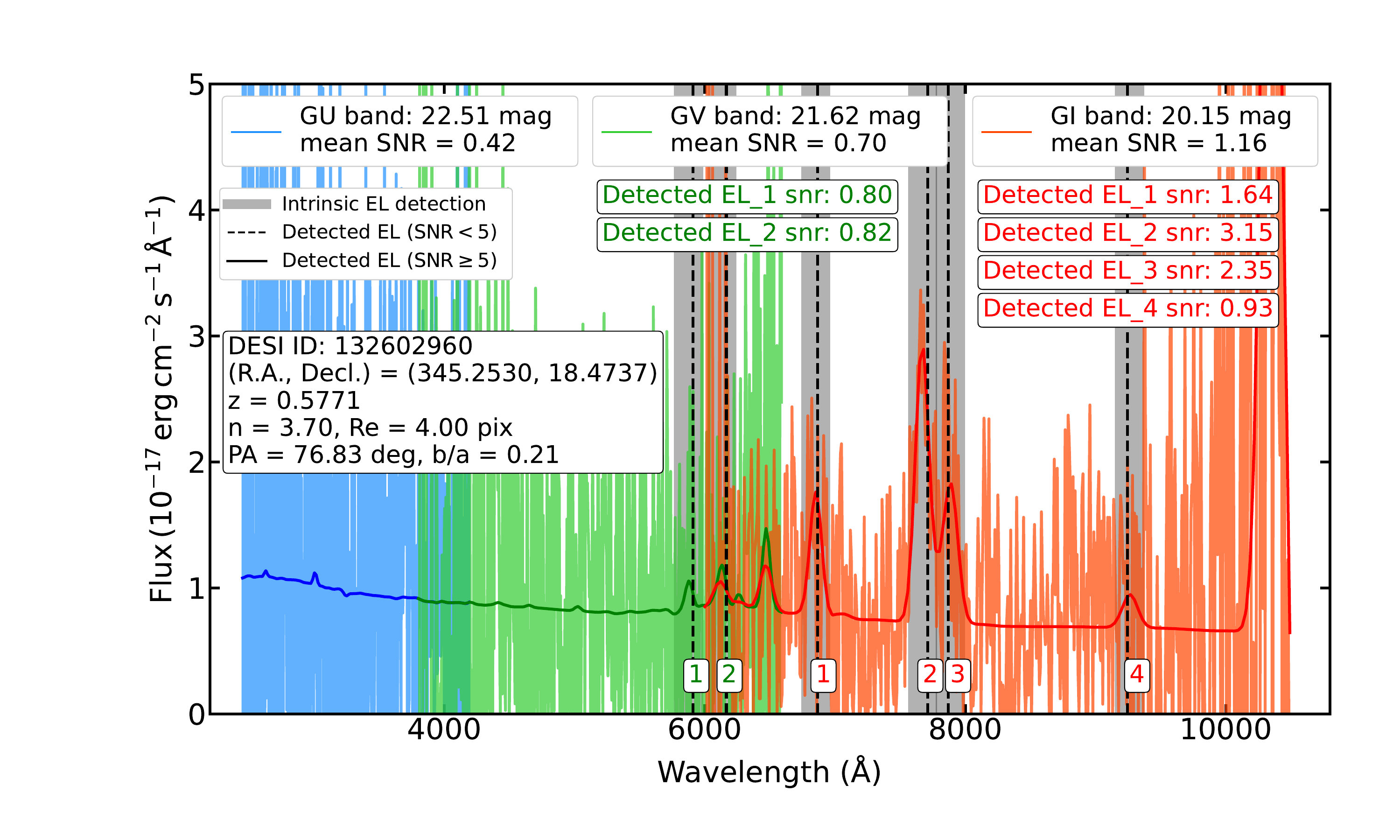}
        \caption{Low mean SNR, low EL SNR}
    \end{subfigure}
    \hfill
    \begin{subfigure}[b]{0.48\textwidth}
        \centering
        \includegraphics[width=\linewidth, angle=0]{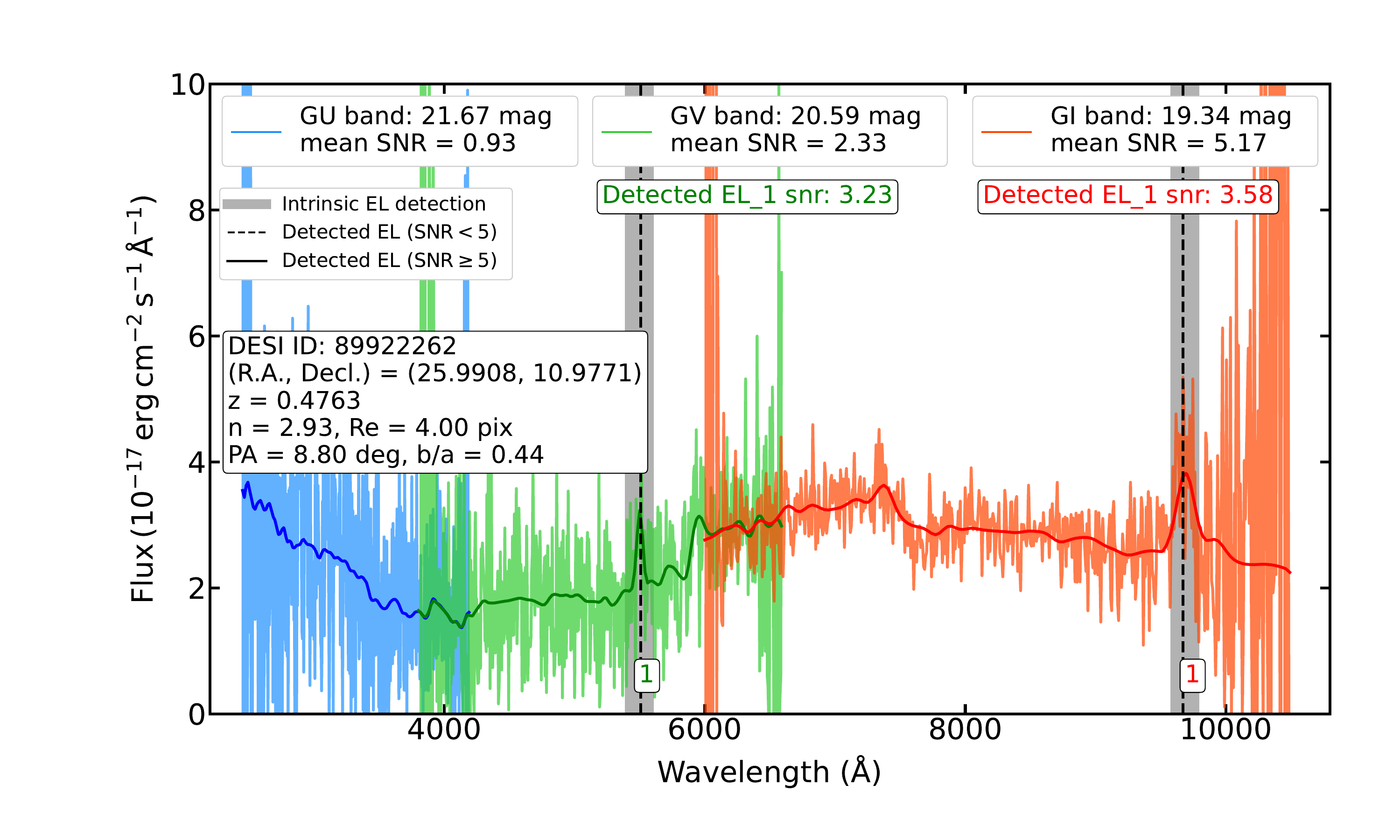}
        \caption{High mean SNR, low EL SNR}
    \end{subfigure}
    \hfill
    \begin{subfigure}[b]{0.48\textwidth}
        \centering
        \includegraphics[width=\linewidth, angle=0]{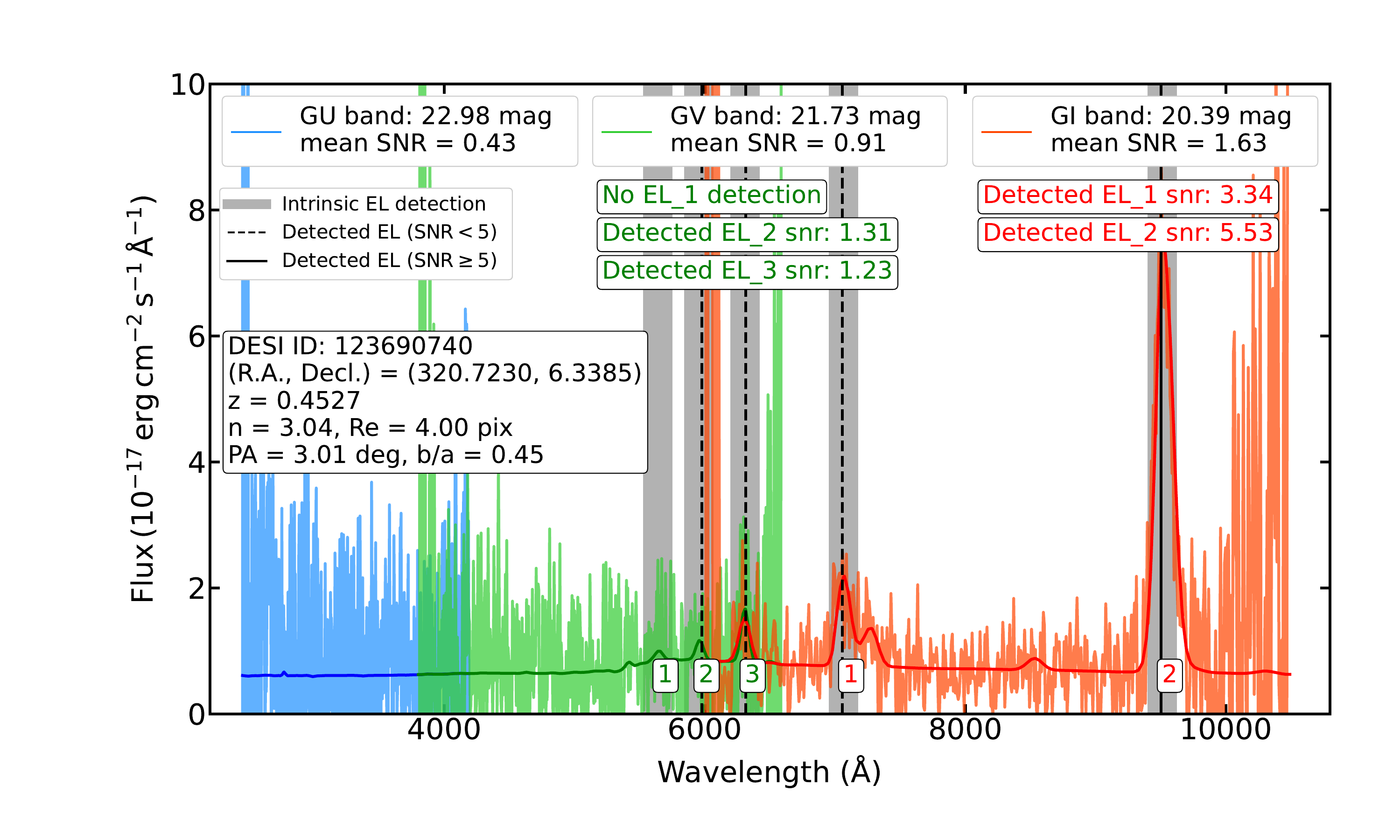}
        \caption{Low mean SNR, high EL SNR}
    \end{subfigure}
    \hfill
    \begin{subfigure}[b]{0.48\textwidth}
        \centering
        \includegraphics[width=\linewidth, angle=0]{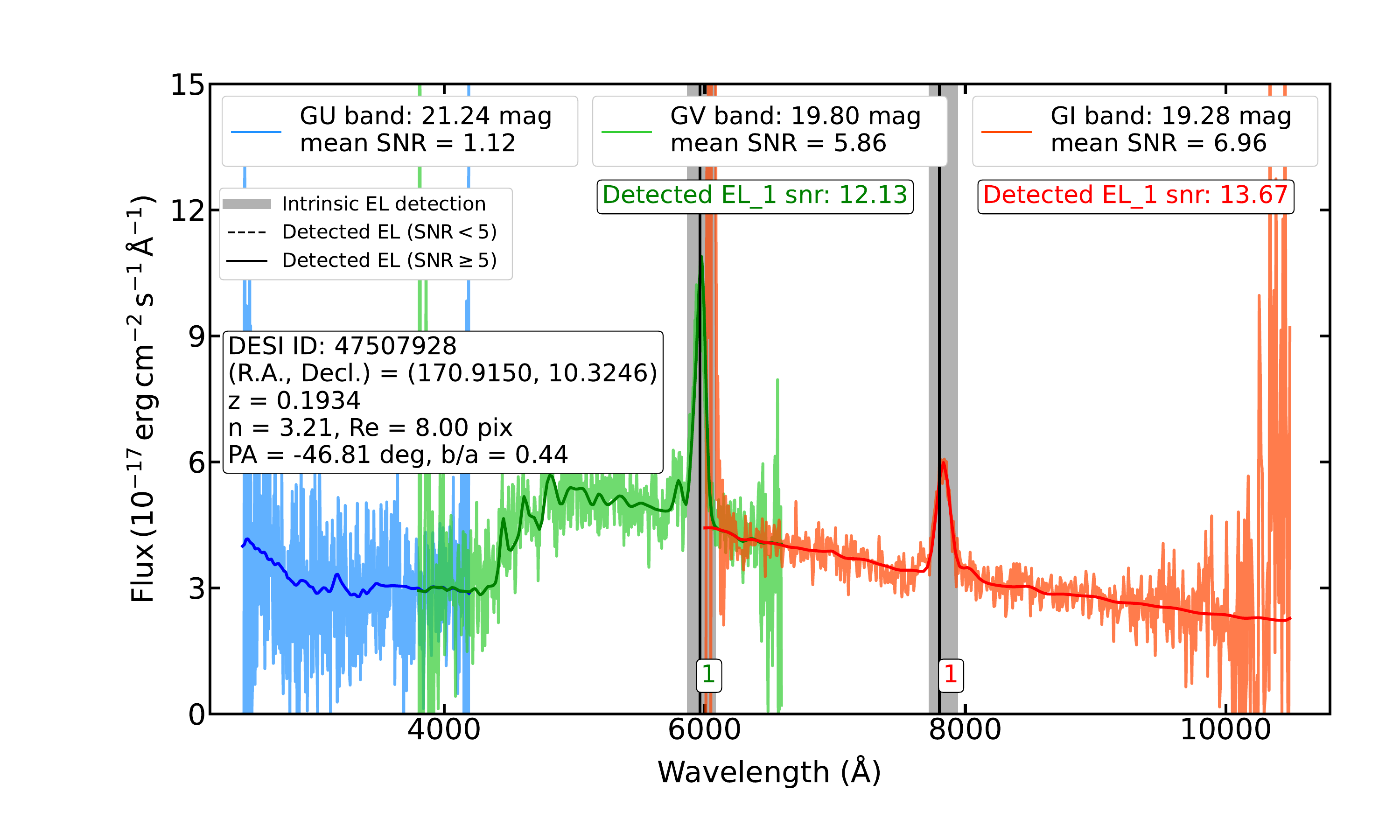}
        \caption{High mean SNR, high EL SNR}
    \end{subfigure}
    \hfill
\caption{ Six examples of CSST slitless spectra simulated by \texttt{CESS}.  In each panel, the simulated signal and signal+noise spectra are presented in three CSST slitless grating bands $GU$ (blue), $GV$ (green), and $GI$ (orange), respectively.  The DESI ID, sky coordinates, redshift, and 2D morphological parameters are labelled on the left.  The CSST grating band magnitudes and mean SNRs are shown at the top.  The grey vertical stripes mark the detected emission lines in the signal spectrum.  The black vertical lines mark the detected emission lines in the signal+noise spectra if they satisfy the successful detection threshold and are in the vicinity of the intrinsic emission lines.  The solid lines mean that they have ${\rm SNR}\geq5$ while the dashed lines refer to ${\rm SNR}<5$. }
\label{fig:emulator_spec_example}
\end{figure*}

\subsection{Spectra overlapping}
\label{sec:overlap}

In slitless spectroscopy, overlap contamination is a common issue that can significantly affect the accuracy of spectral extraction and measurement \citep{Kummel+2009}.  In crowded fields of galaxies (e.g., groups or clusters), the 2D slitless spectra of some galaxies often partially overlap each other.  The contamination from neighbouring sources contributes extra noise to the target's spectrum and increases difficulties and errors in extracting the spectrum.  A specific module is developed to measure the overlap effect in crowded fields and estimate the overlap rate and flux contamination fraction.

Fig.~\ref{fig:overlap} shows the schematic diagram for identifying the cases with overlap contamination.  Over a single 9K\,$\times$\,9K CCD detector, the sky coordinates of input galaxy catalogues are converted into the image coordinates, and the locations and areas of their zero- and first-order slitless 2D spectrum images are identified using the main program of \texttt{CESS}.  This overlap module analyses the locations and areas of the 2D spectra of all input galaxies and identifies the contamination areas and corresponding fluxes from adjacent sources for each of the input target galaxies.

\citet{Treu+2015} presented the contamination fractions in galaxy clusters from the Grism Lens Amplified Survey from Space project.  They proposed three criteria for the mild, moderate, and severe contamination fractions, to quantify the extraction area of the target 2D spectrum overlapped by other sources at a contaminated flux level of $<10$, 10--40, and $>40$\,per\,cent, respectively.  Adopting these criteria, our overlap module is able to calculate the contamination fraction for each of the galaxies in a crowded field.  The overlap rate represents the fraction of galaxies in a crowded field whose 2D spectra are contaminated by other sources.  It describes the degree to which the spectra of all sources in the field are affected by neighbouring sources.  It is worth noting that the contamination levels do not provide a comprehensive description of contamination to a specific source and are only used for reference.  The distribution of contamination fluxes in the simulated 2D spectra of a target is uncertain and complex.  The extraction of the 1D spectrum may not be affected even if its contamination level is severe, such as when the contaminated fluxes are at the edges or ends of the target spectra.

\section{Applications}
\label{sec:applications}

Using \texttt{CESS}, we are able to quickly generate simulated CSST slitless 1D spectra for hundreds of millions of galaxies with high-resolution model spectra.  We apply \texttt{CESS} to a large catalogue of galaxies from the DESI survey and present some statistical results.

\begin{figure}
\centering
\includegraphics[width=1\linewidth, angle=0]{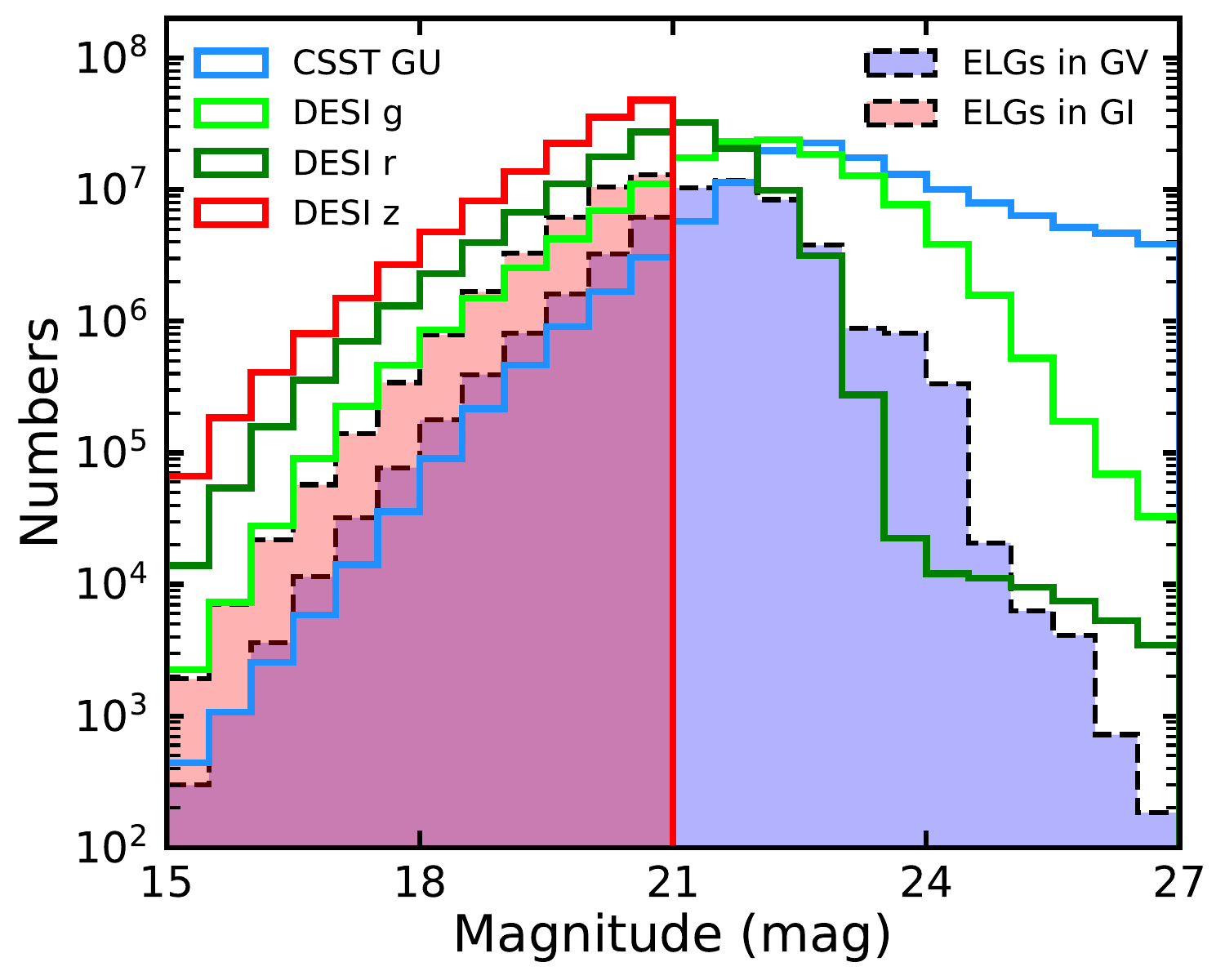}
\caption{The magnitude distribution of all sample galaxies in four bands: $GU$ (light blue) magnitude obtained from \texttt{CESS}, DESI photometric magnitudes in $g$ (light green), $r$ (dark green), and $z$ (red) are presented in hollow histograms.  Galaxies with detected emission lines in $GV$ and $GI$ are presented in blue and pink histograms with dashed black edges, respectively.  }
\label{fig:mag_distribution}
\end{figure}

\subsection{The DESI Galaxy catalogue}
\label{sec:sedlibrary}

In order to assess the cosmological constraining power of the CSST slitless spectroscopic survey, we construct sets of reference mock galaxy redshift surveys (MGRSs\footnote{https://gax.sjtu.edu.cn/data/CSST/CSST.html}) using Jiutian numerical simulations with known cosmological parameters.  The MGRSs are generated via two different approaches, the conditional stellar mass function model from \citet{Yang+2012} and the subhalo abundance matching (SHAM) method from Yang et al. (in preparation).  In this work, we use the SHAM method to generate MGRSs, and more details can be found in Paper I of this series \citep{Gu+2024}.

To properly mimic the observational magnitude selection effect, each galaxy in MGRSs is assigned a DESI $z$-band luminosity using the SHAM method with the galaxies in the DESI Legacy Survey DR9 \citep{Zou+2017,Dey+2019}.  These galaxies in the mock catalogue are kept with redshifts of $z<1.0$ and magnitudes of $m_{z}<21$\,mag.  Multiband photometries as well as images are then assigned to each galaxy using a 3-D parameter space nearest neighbour sampling of the DESI observational galaxies and groups.  To properly model the spectrum for each galaxy, we match each galaxy in our sample with the one in the DESI LS DR9 group catalogue according to its redshift, luminosity and halo mass \citep{Yang+2021}.

To efficiently simulate the 1D CSST slitless spectra, a seed galaxy spectrum library with known parameters is needed.  We make use of all the galaxies in the DESI LS DR9 photometric catalogue that were used in \citet{Yang+2021} for group construction.  This set of galaxies have photometric data in $g$, $r$, $z$, $W1$, and $W2$, with spectroscopic or photometric redshift obtained by \citet{Zhourp+2021} \citep[see also][]{Zou+2019}.  For each galaxy in the seed catalogue, we perform photometric spectrum analysis with \texttt{BayeSED} \citep{Han+2014,Han+2019,Han+2023}.  With the simple stellar population models from \citet{BC+2003}, Chabrier initial mass function \citep{Chabrier+2003}, exp-decline star formation history, \citet{Calzetti+2000} dust extinction law, and nebular emissions modelled with the Cloudy photoionization code \citep{Ferland+2017}, we generate a galaxy spectrum library with known parameters, including best-fitting spectra, redshift, sky coordinates, and physical parameters.  In total, this seed galaxy spectrum library contains 138\,348\,981 galaxies.

\begin{figure*}
\centering
\includegraphics[width=\linewidth]{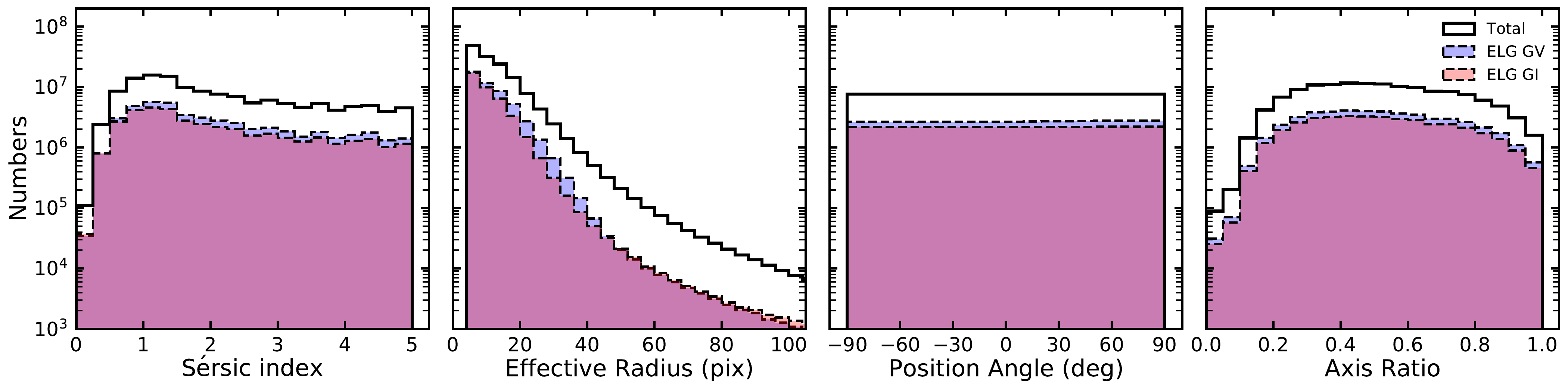}
\caption{ The distribution of 2D morphological parameters of the simulated CSST slitless spectra by \texttt{CESS}. 
\emph{Top left panel}: the distribution of S\'ersic index in range of [0.2, 5]. 
\emph{Top right panel}: the distribution of effective radius in the range of 0--100\,pix. 
\emph{Bottom left panel}: the distribution of position angle in the range of [$-$90, 90$^{\circ}$]. 
\emph{Bottom right panel}: the distribution of axis ratio in the range of [0.1, 1]. 
The histograms of ELGs in $GV$ and $GI$ are shown in blue and pink, respectively. }
\label{fig:2d_morph_distribution}
\end{figure*}

\subsection{Statistical results} 
\label{sec:results}

Utilizing the library of high-resolution spectra as the inputs for \texttt{CESS}, we obtain the simulated CSST slitless 1D spectra.  The content includes the simulated intrinsic signal spectra derived by convolving the input high-resolution spectra with CSST instrument effects, the simulated signal+noise spectra which incorporate various noises, and the SNR spectra for all galaxies in the DESI LS DR9 catalogue.  Furthermore, the detection of emission lines is performed for the simulated signal and signal+noise spectra and the line parameters are also recorded.  Since the major emission lines in the rest-frame optical band mostly appear in $GV$ and $GI$ (see Fig.~\ref{fig:zcoverage}),
the emission line detection is performed in these two bands.  The output catalogues also include relevant parameters, such as the 2D morphological parameters, redshifts, wavelength calibration errors, mean SNRs, and noise RMSs.   The mean SNRs of the whole spectra are estimated by all signals over all noise in the spectral extraction regions.  For emission lines, the SNRs are estimated within their widths related to the morphological parameters (see Fig.~\ref{fig:el_evolution}).

Fig.~\ref{fig:emulator_spec_example} displays six examples of the CSST slitless 1D spectra simulated by \texttt{CESS}.  We show the simulated signal and signal+noise spectra of three CSST grating bands in each panel.  With different emission line detection outcomes, these examples represent six cases: 

\begin{enumerate}
  \item[(a)] No emission lines are detected in both the signal spectrum and the signal\,+\,noise spectrum.
  \item[(b)] Emission lines are detected only in the signal spectrum and not in the signal\,+\,noise spectrum.
  \item[(c)] Emission lines are detected in both the signal spectrum and the signal\,+\,noise spectrum, but the SNRs of the emission lines and the mean SNR of the spectra are low.
  \item[(d)] The mean SNR of spectra is high, but the SNRs of emission lines are low.
  \item[(e)] The mean SNR of spectra is low, but at least one of the emission lines has a high SNR.
  \item[(f)] The mean SNR of spectra and at least one of the emission lines have high SNR. 
\end{enumerate}

\begin{figure*}
\centering
\includegraphics[width=\linewidth]{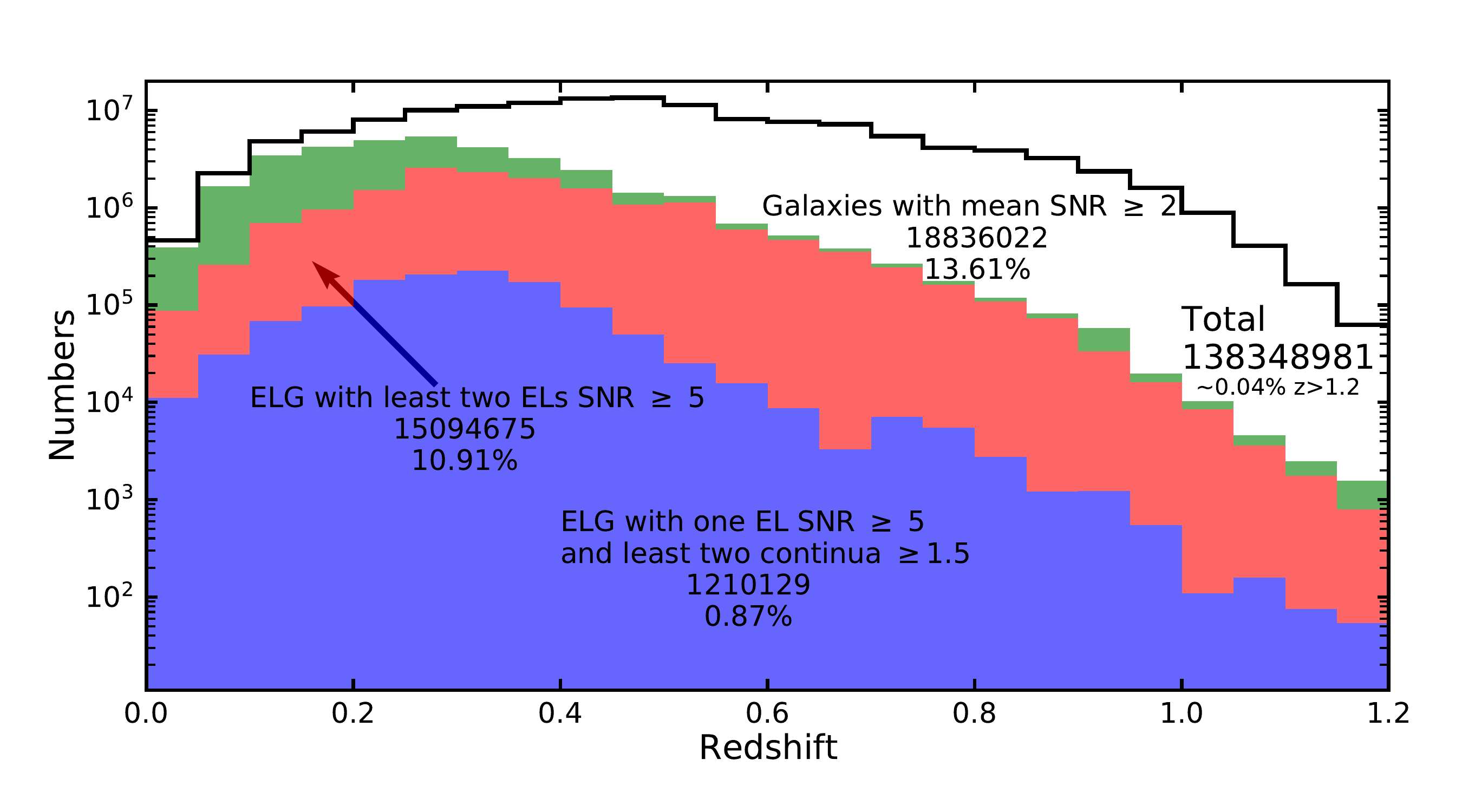}
\caption{ The redshift statistics of \texttt{CESS} simulated galaxies for CSST slitless spectroscopic survey.  The hollow histogram stands for the redshift distribution of the input catalogue.  The colour-coded stacked histograms represent the galaxies selected by the criteria for secure redshift measurements, including ELGs with one emission line of ${\rm SNR}\geq5$ and at least two band continua of ${\rm SNR}\geq1.5$ (blue), ELGs with at least two emission lines of ${\rm SNR}\geq5$ (red), and other galaxies with at least two band continua of ${\rm SNR}\geq2$ (green).   The corresponding numbers and fractions are labelled in the panel.  }
\label{fig:z_distribution}
\end{figure*}

For the (a) and (b) cases, the galaxies are considered to be non-ELGs.  For other cases, galaxies with at least one secure emission line detection, which means the emission line is detected in both signal and signal+noise spectrum and satisfies the successful detection criteria, are then identified as ELGs.

The magnitude distributions of sample galaxies in CSST $GU$, DESI $g$, $r$, and $z$ are presented in Fig.~\ref{fig:mag_distribution}.  The hollow histograms represent all galaxies, and those with emission lines detected are marked by the filled histograms.  The DESI sample galaxies are selected with $m_{z}<21$\,mag.  They are mostly red and faint in $g$ and $GU$.  Galaxies are counted as detections of magnitude $<$ 23.2\,mag in one of the three grating bands.  From the simulated CSST spectra, we estimate that the number of ELGs detected in $GV$ and $GI$ is 48\,686\,871 ($\sim35.19$\,per\,cent) and 39\,518\,878 ($\sim28.57$\,per\,cent), respectively.  The number of ELGs detected in both $GV$ and $GI$ is 25\,146\,356 ($\sim18.18$\,per\,cent).  These results show that the proportion of ELGs identified in at least one band of CSST slitless spectroscopy could reach 45.58\,per\,cent.  The magnitude distributions of ELGs in $GV$ and $GI$ follow these of the entire sample galaxies, as shown in Fig.~\ref{fig:mag_distribution}.

\begin{figure*}
\centering
\includegraphics[width=\linewidth, angle=0]{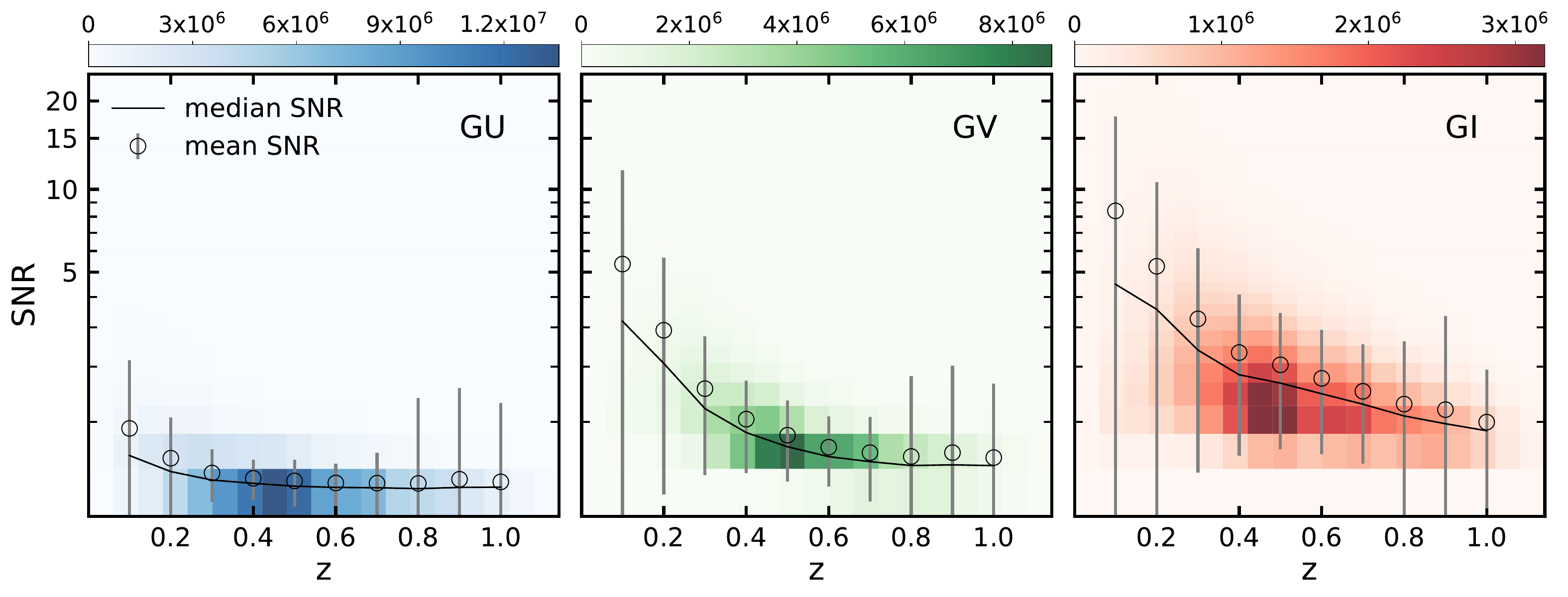}
\caption{The distribution of all sample galaxies in the diagram of spectral SNR versus redshift $z$ for the simulated CSST slitless spectroscopy in $GU$ (left panel), $GV$ (middle panel), and $GI$ (right panel).  The colour-coded greyscale maps show the number densities of sample galaxies.  The black circles are the average SNR in given $z$ bins, and the black solid line draws the median SNR as a function of redshift. }
\label{fig:SNR_z}
\end{figure*}

Distributions of the morphological parameters for all galaxies and ELGs are given in Fig.~\ref{fig:2d_morph_distribution}.  These morphological parameters are empirically assigned to sample galaxies mainly based on the GEMS morphological catalogue.  More than half of the sample galaxies have $n<2.5$.  Their effective radii are mostly $R_{\rm e}\leq15$\,pix, corresponding to about 1\,arcsec, indicating that most of the input galaxies are compact.  Their slitless spectroscopy is marginally affected by the self-broadening effect.  The majority of sample galaxies have an axis ratio of $0.3<b/a<0.8$.  The fraction of nearly edge-on galaxies with $b/a<0.2$ is less than 5\,per\,cent.  The position angles are uniformly distributed and do not affect the actual selection of the 2D brightness profiles.  The distributions of ELGs in $GV$ and $GI$ are also consistent with the overall sample.

\subsection{Redshift completeness} 
\label{sec:redshifts}

Aiming at evaluating the outputs for the CSST redshift survey, we analyse the simulated CSST spectra and identify their spectral features for secure redshift measurements.  Based on emission line detections in $GV$ and $GI$, the sample galaxies can be divided into three groups: 15\,367\,074 ($\sim11.11$\,per\,cent) galaxies with only one emission line detected, 47\,692\,319 ($\sim34.47$\,per\,cent) galaxies with more than one emission line detected, and 75\,289\,588 ($\sim54.42$\,per\,cent) galaxies with no emission lines detected.

Secure redshifts can be measured from either emission lines or absorption lines reliably detected in spectra.  The detection of absorption lines requires the continuum at a high SNR level.  We propose three criteria for secure redshift measurements: (1) two or more emission lines of ${\rm SNR}\geq5$; (2) a single emission line with ${\rm SNR}\geq5$ and continua in $GV$ and $GI$ have a mean ${\rm SNR}>1.5$; (3) continua in $GV$ and $GI$ have a mean ${\rm SNR}\geq2$ for non-ELGs.

Fig.~\ref{fig:z_distribution} displays the redshift distribution of the DESI galaxies in the sample and the simulated galaxies with spectral features at given detection levels.  The black solid histogram stands for the total sample in $z<1.2$, with a fraction of 99.96\,per\,cent.  In total, about 25.39\,per\,cent galaxies are expected to have reliable redshift measurements from the CSST slitless redshift survey.  The ELGs with only one emission line detected at ${\rm SNR}\geq5$ represent 0.87\,per\,cent of the total sample.  We notice a small bump at $z=0.7$, likely attributed to the \OII\ line shifted into $GI$.  The single emission line criterion also drops suddenly at $z>1$, where the \Hb\ and \OIII\ shift out of $GI$ (see Fig.~\ref{fig:zcoverage}).  The ELGs with at least two emission lines detected with ${\rm SNR}\geq5$ represent 10.91\,per\,cent of the total sample.  They are mostly distributed at around $z\sim0.3$, where \OIII\ and \Ha\ lines can be observed in $GV$ and $GI$, respectively.  The non-ELGs with reliable continua represent 13.61\,per\,cent of the total sample.  The majority of these galaxies are located at $z<0.4$.

In Fig.~\ref{fig:SNR_z}, we show the mean SNR as a function of redshift for all sample galaxies in $GU$, $GV$, and $GI$.  The mean SNR of the simulated spectra are shown in each panel, while a colour bar represents the number of galaxies in the grid.  The medians of the mean SNRs in given redshift bins are also denoted.  The majority of sample galaxies have a mean ${\rm SNR}<1$ in $GU$, suggesting that their $GU$ slitless spectra are barely detectable.  The low-$z$ galaxies tend to have high SNR in $GV$ and $GI$, and the mean SNR drops below two for the bulk of galaxies at $z>0.5$.  Importantly, our calculation shows that DESI galaxies are best detected in $GI$, having ${\rm SNR}>\sim1$ for most of them.

In summary, the availability of comprehensive emission line data, including emission line detections, SNRs, and other relevant parameters, will greatly facilitate the assessment of CSST capabilities for redshift measurements and cosmological studies.

\begin{figure}
\centering
\includegraphics[width=\linewidth, angle=0]{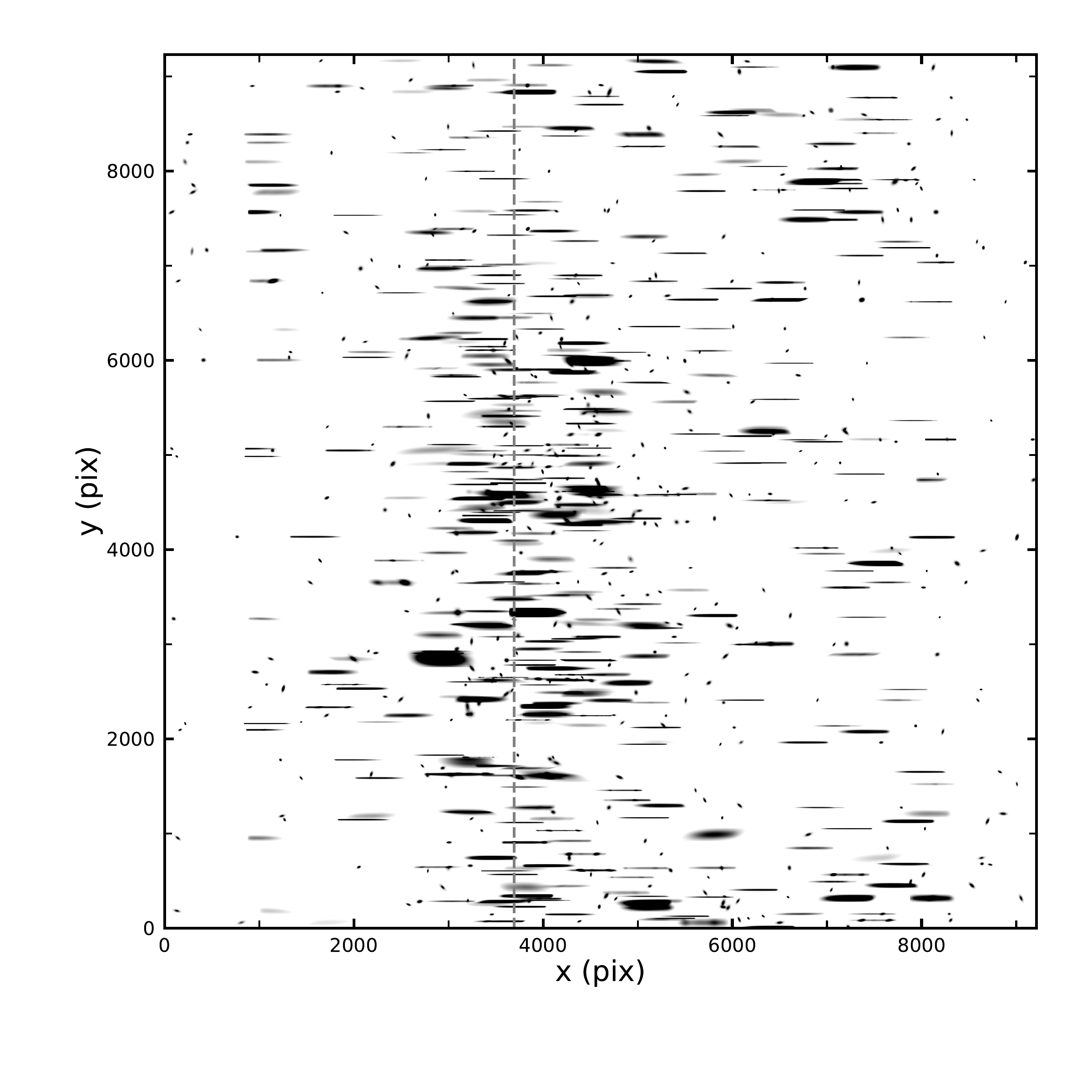}
\caption{ An example of the simulated zero- and first-order slitless image (without noise) of a DESI cluster on a 9K\,$\times$\,9K CCD by \texttt{CESS}.  The DESI cluster has a richness of 119.75 with its density peak at $\alpha = 02:59:04.60$, $\delta = -06:59:27.83$ \citep{Zou+2021}.  This image contains 438 galaxies from the DESI catalogue, including the member galaxies of the cluster, and the foreground and background galaxies.  The grey-dashed line presents the dispersion boundary of the two parts of the CSST grating in front of the detector.  }
\label{fig:contamination_image}
\end{figure}

\subsection{Completeness in galaxy clusters}
\label{sec:completeness}

We adopt the DESI cluster catalogue from \citet{Zou+2021} to estimate the overlap effect as a function of cluster richness.  This cluster catalogue includes 540,432 galaxy clusters at $z\lesssim1$ identified in the DESI LS DR9 catalogue with their sky coordinates and richness.  Fig.~\ref{fig:contamination_image} shows an example of the simulated slitless spectra image for a DESI cluster.  Using the contamination estimation module described in Section~\ref{sec:overlap}, galaxies covered by this single CCD are dispersed into zero-order images and first-order spectra.  The contamination fractions are calculated based on such images.

\begin{figure}
\centering
\includegraphics[width=\linewidth, angle=0]{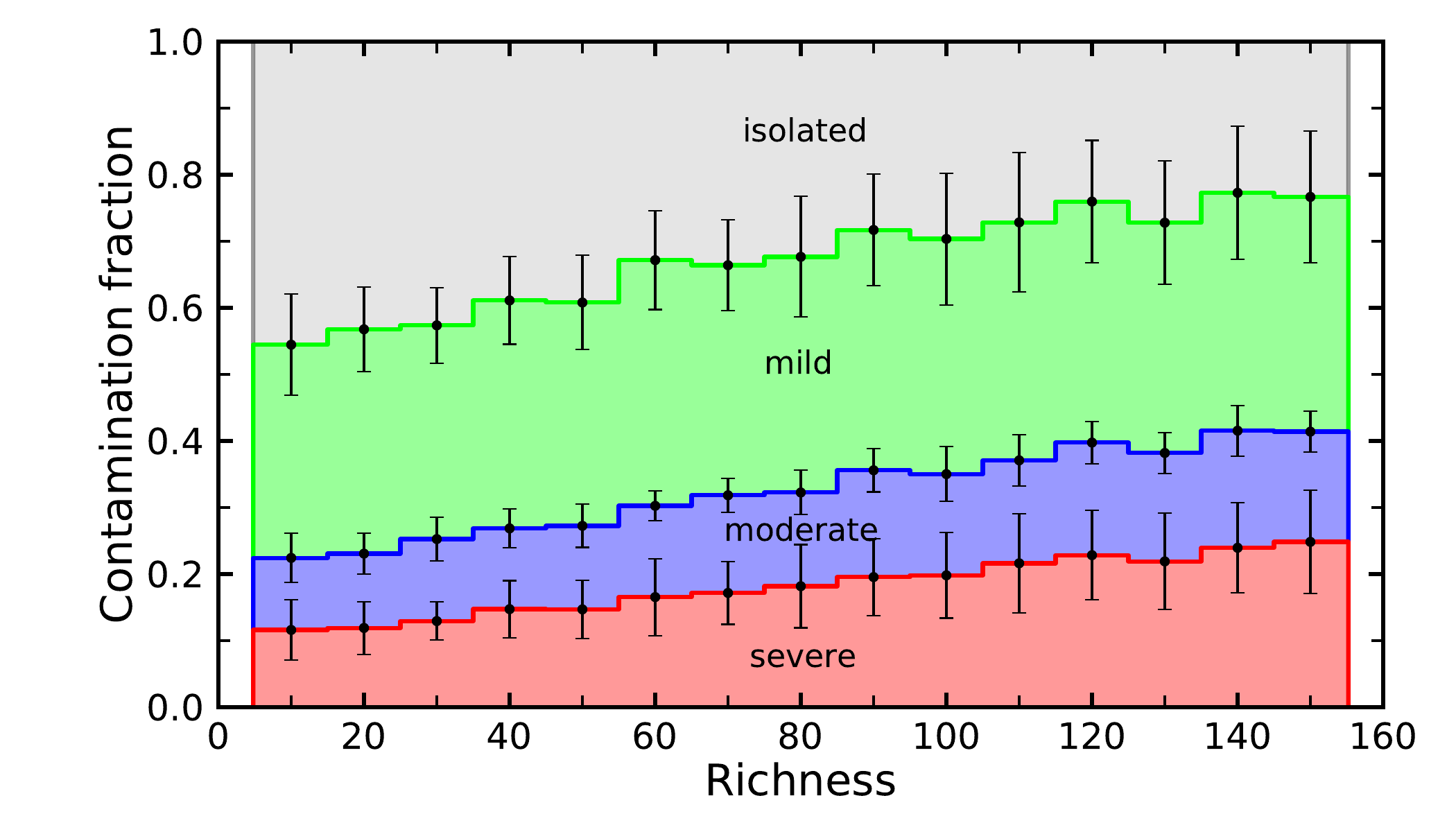}
\caption{ The contamination fraction of CSST slitless spectra in the DESI clusters with different richnesses.  Each richness bin contains 100 clusters randomly selected from \citet{Zou+2021}.  The mean values and standard deviations of each bin are marked as black error bars.  The fractions are shown as a function of cluster richness for contamination at the level of mild ($<10$\,per\,cent, green), moderate (10--40\,per\,cent, blue), and severe ($>40$\,per\,cent, red), respectively.  The isolated galaxies are labelled as grey.  }
\label{fig:contamination}
\end{figure}

In total, a sample of 1500 clusters evenly spanning into 15 richness bins ranging from 10 to 150 is used to investigate the contamination fraction across richness, as shown in Fig.~\ref{fig:contamination}.  The isolated fraction represents the galaxies with no contamination fluxes estimated in their first-order slitless spectra.  While the overlap rate is defined as the summary of the three contamination fractions.  The overlap rate rises from 0.55 to 0.8 at the richness from 10 to 150.  In detail, the contamination fraction increases from 0.1 to 0.25 for the severe case, from 0.1 to 0.15 for the moderate case, and remains nearly constant at about 0.35 for the mild case.  The variation of the severe case is larger than the mild and moderate cases, which mostly contributes to the overlap rate in the high richness clusters.

Note that the estimated overlap rate and contamination fractions estimated here only count galaxies in the DESI LS DR9 catalogue for each cluster field.  We also only consider the zero- and first-order slitless image of each galaxy.  When taking into account contamination from other orders of slitless spectra, stellar sources, and the intergalactic light from the centre of the galaxy clusters, the contamination fractions and overlap rate could become even higher.   Moreover, as the vast majority of the galaxies in our seed catalogue are located in poor groups \citep{Yang+2021}, we will perform a more detailed analysis of this effect in the following work by considering groups with, for example, 1, 2, 4, and 8 members.  

\section{Conclusions} 
\label{sec:conclusion}

Aiming at testing the reliability of spectroscopic redshift measurements for the slitless spectroscopic redshift survey with the upcoming CSST, we developed an emulator \texttt{CESS} to simulate CSST slitless 1D spectra for hundreds of millions of galaxies with limited computing resources.  Empirical relations of observed galaxies and morphological catalogues are used to quantify the instrumental and morphological-broadening effects and quickly generate mock CSST slitless 1D spectra.  Our program \texttt{CESS} allows one to generate the intrinsic signal spectra that match the CSST instrumental parameters, and the observed signal+noise mock spectra that account for the morphological broadening effects through empirically assigned 2D brightness profiles for the input galaxies of known redshift, stellar mass, magnitude, and spectral type.

We quantified the self-broadening effects on the simulated 1D slitless spectra by the 2D brightness profiles of galaxies, as functions of morphological parameters of S\'ersic index ($n$), effective radius ($R_{\rm e}$), position angle (PA), and axis ratio ($b/a$).  The changes in FWHM and peak intensity of emission lines are used to measure the broadening effect.   We found that $n$ has little or no effect on broadening the spectral features.  In contrast, the half-light radius is the dominant morphological parameter influencing the broadening of the slitless spectroscopy and the impact begins to be significant at $R_{\rm e}\leq$1.1\,arcsec.  Axis ratio and position angle alone play a secondary role but may magnify the broadening effect when the galaxy is close to edge-on and the effective radius is relatively large ($R_{\rm e}>$2.3\,arcsec).

The emulator \texttt{CESS} provides two ancillary modules.  One is for the detection of emission lines in the simulated slitless 1D spectra and measures the central wavelength, total flux, and SNR of these emission lines.  
The other is to evaluate the contamination fraction in crowded fields such as groups and clusters so that a better estimate of redshift completeness can be conducted for the CSST redshift survey.

We applied \texttt{CESS} to a sample of 138\,348\,981 galaxies selected from the DESI LS DR9 catalogue with known galaxy parameters and high-resolution model spectra best fitting to their broad-band SEDs and successfully obtained the simulated CSST slitless 1D spectra.  The SNR distribution as a function of redshift is also estimated in three CSST slitless grating bands.  The $GU$ slitless spectra have SNR below 1 in all redshift bins, while the $GI$ has the best SNR distribution among the three bands with ${\rm SNR}>\sim1$ for most of them.

Our results show that galaxies are detected by CSST slitless spectroscopy with at least one band $<$ 23.2\,mag.  About 35.19 and 28.57\,per\,cent of ELGs in the sample could be spectroscopically detected in $GV$ and $GI$, respectively.  The fraction of ELGs with a single detected line is about 11.11\,per\,cent, and 34.47\,per\,cent of ELGs have multiple emission lines securely detected.  For non-ELGs, the fraction is 54.42\,per\,cent. 
We measured SNRs for emission lines and continua and identified reliable detections for secure redshift measurements.  Of the sample, 10.91\,per\,cent have at least two emission lines detected at ${\rm SNR}>5$, 0.87\,per\,cent have only one emission line of ${\rm SNR}>5$ and continua with ${\rm SNR}>1.5$ in GV and GI, and 13.61\,per\,cent have continua with ${\rm SNR}>2$ in GV and GI.

The morphologies of our results fitted from the GEMS catalogue show that the majority of the output galaxies are late-type and compact galaxies with a moderate axis ratio.  The self-broadening effect is negligible in the slitless spectroscopy for these galaxies.  Furthermore, the contamination fractions evaluated in DESI cluster fields provide a lower limit for the overlap rate and contamination level with different richness.  At increasing richness, the overlap rate increases from 0.55 to 0.8, with the moderate and severe fractions increasing from 0.1 to 0.15 and from 0.1 to 0.25, respectively.  The mild fraction remains constant at about 0.35.  The contamination levels serve as a reference for that the distribution of contaminated fluxes is complex for a specific source.  In addition, these fractions might be underestimated because the noise level is underestimated in the central regime in a grating CCD detector by \texttt{CESS}.

We caution that our calculations in \texttt{CESS} might slightly overestimate the SNR for slitless spectra because of simplifying the complex instrumental effects.  Our work does not consider the secure calibration of wavelength and flux.  In addition, the higher orders of slitless spectra which are not considered in \texttt{CESS} may contribute more contamination fluxes in clusters.  The PSF variations across the large FOV of CSST are also ignored.  \texttt{CESS} can be improved by mimicking the 1D slitless spectra more realistic for CSST.

In future works, redshift measurements and completeness analyses will be carried out using the simulated spectra of CSST slitless spectroscopy based on the DESI galaxy catalogue and \texttt{CESS}, and quantitatively evaluate the impact on large-scale structure power spectrum measurement.

\section*{Acknowledgements}

We are grateful to the anonymous referee for his/her valuable comments and suggestions that helped to improve this manuscript. 
We thank Zhongxu Zhai and Fengwu Sun for the helpful discussions and some colleagues who provided helpful suggestions for this work. 
The computations in this paper were run on the Starburst at Purple Mountain Observatory.  
This work is supported by the science research grants from the China Manned Space Project with NO. CMS-CSST-2021-A02, CMS-CSST-2021-A04, CMS-CSST-2021-A06 and CMS-CSST-2021-A07, the National Key Research and Development Program of China (2023YFA1608100) and the National Science Foundation of China (12288102, 12233005, 12073078, 11890692, 11833005 and 11773063).

\section*{Data Availability}

The data underlying this paper will be shared on reasonable request to the corresponding author.



\bibliographystyle{mnras}
\bibliography{cess_mnras} 

\bsp	
\label{lastpage}
\end{document}